\documentclass[acmsmall,screen,authorversion,nonacm]{acmart} % use this for arxiv

\usepackage{multirow}
\usepackage{algorithmic}
\usepackage[linesnumbered,ruled,vlined]{algorithm2e}
\usepackage{array}
\usepackage{booktabs}
\usepackage{color}
\usepackage{graphics}
\usepackage{adjustbox}
\usepackage{longtable}
\usepackage{hhline}
\usepackage{subcaption}
\usepackage{hyperref}
\usepackage{bbding}
\usepackage{makecell}
\usepackage{tikz, pgfplots}
\usepackage{tablefootnote}
\usepackage{threeparttable}
\usetikzlibrary{positioning}
\usepackage{pgf-pie} 

\AtBeginDocument{%
  }

\setcopyright{none} % use none for arXiv, acmcopyright for ACM
\acmDOI{XXXXXXX.XXXXXXX}

\begin{document}

\title{An Overview of FPGA-inspired Obfuscation Techniques}

\author{Zain Ul Abideen}
\email{zain.abideen@taltech.ee}
\orcid{0000-0002-2395-0178}
%\email{webmaster@marysville-ohio.com}
\affiliation{%
  \institution{Centre for Hardware Security, Tallinn University of Technology}
  \streetaddress{Akadeemia tee 15a}
  \city{Tallinn}
  \state{Harju}
  \country{Estonia}
  \postcode{12611}
}

\author{Sumathi Gokulanathan}
\email{sumathi.gokulanathan@taltech.ee}
\orcid{0000-0002-7065-4400}
\affiliation{%
  \institution{Centre for Hardware Security, Tallinn University of Technology}
  %\streetaddress{1 Th{\o}rv{\"a}ld Circle}
  \city{Tallinn}
  \country{Estonia}}

\author{Muayad J. Aljafar}
\email{muayad.al-jafar@taltech.ee}
\orcid{0000-0001-7336-4444}
\affiliation{%
 \institution{Centre for Hardware Security, Tallinn University of Technology}
 \streetaddress{Akadeemia tee 15a}
 \city{Tallinn}
 \state{Harju}
 \country{Estonia}
  \postcode{12611}}

\author{Samuel Pagliarini}
\email{samuel.pagliarini@taltech.ee}
\orcid{0000-0002-5294-0606}
\affiliation{%
  \institution{Centre for Hardware Security, Tallinn University of Technology}
  \streetaddress{Akadeemia tee 15a}
  \city{Tallinn}
  \state{Harju}
  \country{Estonia}
  \postcode{12611}}

\thanks{This work has been conducted in the project ``ICT programme'' which was supported by the European Union through the ESF}
\renewcommand{\shortauthors}{Zain Ul Abideen et al.}

\begin{abstract}
Building and maintaining a silicon foundry is a costly endeavor that requires substantial financial investment. From this scenario, the semiconductor business has largely shifted to a fabless model where the Integrated Circuit supply chain is globalized but potentially untrusted. In recent years, several hardware obfuscation techniques have emerged to thwart hardware security threats related to untrusted IC fabrication. Reconfigurable-based obfuscation schemes have shown great promise of security against state-of-the-art attacks -- these are techniques that rely on the transformation of static logic configurable elements such as Look Up Tables (LUTs). This survey provides a comprehensive analysis of reconfigurable-based obfuscation techniques, evaluating their overheads and enumerating their effectiveness against all known attacks. The techniques are also classified based on different factors, including the technology used, element type, and IP type. Additionally, we present a discussion on the advantages of reconfigurable-based obfuscation techniques when compared to Logic Locking techniques and the challenges associated with evaluating these techniques on hardware, primarily due to the lack of tapeouts. The survey's findings are essential for researchers interested in hardware obfuscation and future trends in this area. 
\end{abstract}

\ccsdesc[500]{Security and privacy~Systems security}
\ccsdesc[500]{Security and privacy~Hardware security implementation}
\ccsdesc[500]{Security and privacy~Hardware reverse engineering}
\ccsdesc[500]{Security and privacy~Hardware attacks and countermeasures}
\ccsdesc[500]{Hardware~Programmable logic elements}
\ccsdesc[500]{Hardware~Programmable interconnect}
\ccsdesc[500]{Hardware~Physical design (EDA)}
\ccsdesc[500]{Hardware~Application-specific VLSI designs}
\ccsdesc[500]{Hardware~Application specific integrated circuits}
\ccsdesc[500]{Hardware~Analysis and design of emerging devices and systems}

\keywords{Hardware security, Trustworthy hardware, Logic obfuscation, FPGA redaction, reconfigurable logic, LUT-based obfuscation}

\received{25 May 2023}
%\received[revised]{12 March 2023}
%\received[accepted]{5 June 2023}

\maketitle

%%%%%%%%%%%%%%%%%%%%%%%%%%%%%%%%%%%%%%%%%%%%%%%%%%%%%%%%%%%%%%%%%%%%%%%%%%%%%%%
%% SECTION - 1 (Introduction)
%%%%%%%%%%%%%%%%%%%%%%%%%%%%%%%%%%%%%%%%%%%%%%%%%%%%%%%%%%%%%%%%%%%%%%%%%%%%%%%
\section{Introduction} \label{sec:Intro}
Integrated Circuit (IC)-based systems have been used in both consumer and military electronics for several decades, enabling a range of devices, from smartphones to satellites. The continued advancements in technology have also led to the adoption of IC-based systems in newer domains like the Internet of Things (IoT) and multi-cloud environments \cite{chip_app}. %ICs are essential components of these electronic devices, playing a crucial role in providing efficiency and performance for various applications. IC-based systems aim at making human tasks even more efficient, convenient, and faster. 
In every domain, the demand for high-performance ICs is increasing. The reason behind this trend is the complexity of modern systems and the need for faster speeds to handle larger amounts of data being processed. As a result, the semiconductor industry is experiencing a surge in demand for products such as memory chips, microprocessors, and sensors. For example, the global IC market is forecast to grow from \$489 billion in 2021 to \$1.136 trillion in 2028 \cite{numberDevices}. On the other hand, ICs require advanced manufacturing processes and specialized equipment, which are only available in a limited number of foundries. %Building an advanced foundry, especially in the semiconductor industry, poses significant financial and technological challenges.

As the industry continues to evolve, the complexity of building and maintaining a foundry increases, resulting in skyrocketing costs. As an example, the estimated cost of building a 3nm foundry is in the range of \$15-20B \cite{Cost3nm}. Thus, contemporary semiconductor vendors are increasingly adopting a \emph{fabless model}, where a globalized IC supply chain allows the production of high-performance ICs without the requirement of heavy investment in specialized foundry equipment. %that can compete on a global scale. %An example of this challenge is Apple's continuing to outsource the manufacturing of their processor chips to TSMC\cite{AppleTsmc}. 

\begin{figure*}[tb!]
    \centerline{\includegraphics[width=1.0\linewidth]{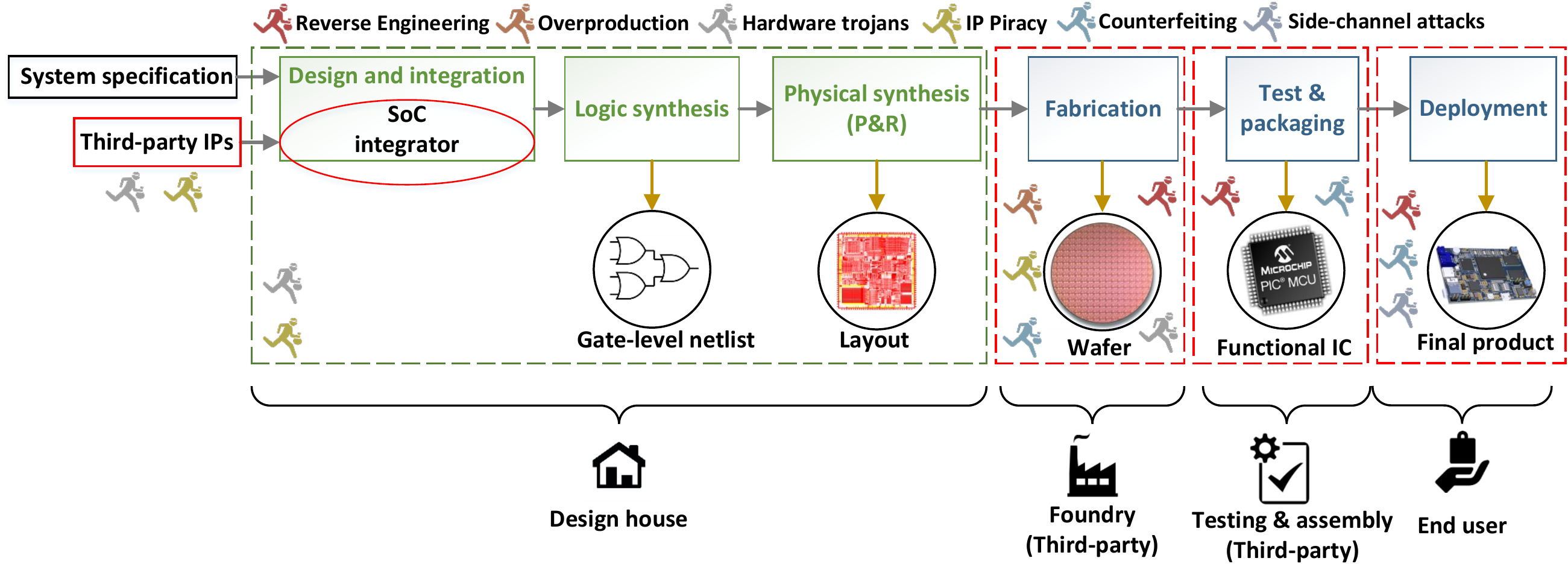}}
    \caption{Typical stages involved in the IC design and in the supply chain, untrusted stages are shown in red.}
    \label{fig:design_flow}
\end{figure*}

The globalized IC supply chain enables design houses to have access to high-end semiconductor manufacturers \cite{Rostami2014}, but the exposure of the layout design to untrusted entities poses significant security threats as shown in Fig. \ref{fig:design_flow}. Losses due to security threats could be severe, including service interruption, damage to public data integrity, monetary losses, etc. For instance, the EU and US warned about the dangers to national security that scammers exploited in the recent IC supply chain crunch \cite{counterfeit_2022}. Similarly, the International Telecommunication Union (ITU) and the European Union Intellectual Property Office (EUIPO) reported in 2015 that 12.9\% of the total sales of smartphones were lost due to counterfeit electronics. The sales of counterfeit devices in the market caused a loss of EUR 45.3 billion to legitimate industries -- a significant monetary loss. The green color in Fig. \ref{fig:design_flow} illustrates the steps performed in a trusted environment. In practice, all involved parties provide assurances but cannot provide guarantees related to the integrity and trustworthiness of the ICs. This lack of guarantees is primarily due to the concept of zero-trust in which one must assume that the foundry and its employees are potential adversaries. The fabrication phase holds the utmost importance as the foundry has access to all low-level details of the design. The related security threats include reverse engineering, overproduction, insertion of hardware Trojans, IP piracy, and counterfeiting \cite{ref_2_nist}.

%%%%%%%%%%%%%%%%%%%%%%%%%%%%%%%%%%%%%%%%%%%%%%%%%%%%%%%%%%%%%%%%%%%%%%%%%%%%%%%
%% FIGURE 
%%%%%%%%%%%%%%%%%%%%%%%%%%%%%%%%%%%%%%%%%%%%%%%%%%%%%%%%%%%%%%%%%%%%%%%%%%%%%%%

\textbf{Reverse engineering} is extensively demonstrated in the literature, as a method to extract the design and/or technology details of an IC with the help of tools and imaging techniques. It involves a complex process of removing the package of an IC, extracting all the layers, stitching the individual layers, and analyzing the obtained images to recover the netlist of a design \cite{RE_1}. %Apart from this, the goal of reverse engineering could also be to insert malicious logic in the design. The adversary might be looking for potential places to insert malicious logic. However, this is not always true because the adversary might exploit engineering change order (ECO) - a post-layout modification as demonstrated in \cite{trojan_tiago}. 
Reverse engineering also provides an opportunity for IP piracy and counterfeiting. Reverse engineering could be exploited in conjunction with other techniques to extract secret information, i.e., cryptographic keys. Reverse engineering becomes more onerous and time-intensive after fabrication, packaging, or even deployment has occurred, but a skilled adversary can still perform it.

As stated earlier, third-party entities are involved in the design, packaging, and testing processes. The design process also typically includes outsourcing third-party IPs, \textbf{piracy} may be evident in different degrees in the form of IP theft, overbuilding at an untrusted foundry, or illegal ownership claims. An unauthorized individual inside the foundry has the potential to steal information through reverse engineering or unlawfully sell the IPs without the authorization of the owner. On the other hand, the untrusted foundry could also be interested to overproduce ICs and sell them in the grey (or black market) at cheaper prices. This is possible because a foundry typically incurs only a marginal increase in costs when manufacturing additional ICs from the same masks \cite{overbuilding_ref}, i.e., the design house that owns the IP bares all the design-related NRE costs . %This may result in the foundry selling the ICs at a lower price than the original IC company. In some cases, overbuilding may not be as cost-effective for trusted foundries, as they already have higher quality standards and are more reliable .

%%IC Camouflaging is a layout-level technique, where the function of selected gates is somehow hidden. Logic Locking is a circuit-level technique to hide the functionality of a design behind a secret key. Split-Manufacturing is a fabrication-level technique that splits a design into trusted and untrusted planes, to be fabricated at trusted and untrusted foundries, respectively. Unfortunately, none of these techniques is currently adopted in large scale production of ICs, for reasons that include (lack of) practicality \cite{split_1} and insufficient security guarantees \cite{eval_logic}

In particular, \textbf{hardware Trojans} are modifications in the form of small and hard-to-detect logic for malicious purposes. Hardware Trojans are utilized to interrupt the service of an IC \cite{hardware_trojan1} or to extract secret information \cite{trojan_tiago}. In the context of the layout, the footprint of the Trojan could be very small which might become invisible to identify and test, especially when there is no reference (golden) design available to cross-verify the functionality. The source of the malicious logic could be a third-party IP or it could also be mounted during manufacturing. As mentioned earlier, the foundry has complete access to the layout therefore it can easily identify potential locations for Trojan insertion \cite{trojan_tiago}. The Trojans/backdoors in third-party IPs may also contain hidden functionalities to expose restricted parts of the design and/or extract some secret information. 

\textbf{Counterfeit} ICs are fraudulently made in such a way to appear almost identical to the original ICs. Counterfeit ICs are divided into seven different classes: recycled, remarked, overproduced, out-of-spec/defective, cloned, forged documentation, and tampered (See Fig. 3 in \cite{ref_5_counterfeit}). Recycled, remarked, out-of-spec/defective, and forged documentation are post-fabrication issues that appear when the counterfeit ICs deployed in a product are outsourced from non-authorized vendors or duplicate IC sellers. On the contrary, overproducing, cloning, and tampering are fabrication-time issues that could be completely tackled if (and only if) all the stages in Fig. \ref{fig:design_flow} were fully executed in a trusted environment. Recalling again, the design house has to share the layout of the design thus it exposes all the minor details to untrusted foundries. On the other hand, plenty of techniques have been developed as countermeasures to these threats. This process is evolving with time and there is a race to bring a novel technique that is resistant and practical \cite{Engels2022}.

Countermeasure techniques to increase the IC security include Logic Locking \cite{logic_1, logic_2, logic_3, logic_4, logic_5}, Camouflaging \cite{cam_1, cam_2, cam_3}, Split Manufacturing \cite{split_1, split_2}, and reconfigurable-based obfuscation techniques \cite{reconfigure_1, reconfigure_3, eASIC, kolhe2022silicon, lut1, Zain_TCAD, chowdhury2021enhancing, lut2, lut3, kolhe2021securing, lut5, attaran2018static, winograd2016hybrid, yang2018exploiting, e-FPGA1, e-FPGA2, e-FPGA3, trap, e-FPGA4, mohan2021hardware, chen2021area, patnaik2018advancing, intro_epic, tomajoli2022alice, rangarajan2020opening, sathe2022investigating}. Reconfigurable-based techniques typically draw inspiration from Field-Programmable Gate Array (FPGA) devices, thus the title of this article. In general, the aforementioned techniques aim to provide security against the threats that occur during IC fabrication. Some techniques also offer degrees of protection for post-fabrication attacks. %Generally speaking, the aforementioned techniques have been proposed to prevent these threats. But, none of these solutions directly address the trustworthiness in the fabrication process \cite{eval_logic}. In summary, there is an urgent need for novel techniques that offer \emph{practical security of ICs} against these threats.But, a new obfuscation technique utilizing embedded-field programmable gate array (eFPGA) or reconfigurable elements has emerged in recent years \cite{reconfigure_1, reconfigure_3, eASIC, kolhe2022silicon, lut1, Zain_TCAD, chowdhury2021enhancing, lut2, lut3, kolhe2021securing, lut5, attaran2018static, winograd2016hybrid, yang2018exploiting, e-FPGA1, e-FPGA2, e-FPGA3, trap, e-FPGA4, mohan2021hardware, chen2021area, patnaik2018advancing, intro_epic, tomajoli2022alice, rangarajan2020opening, sathe2022investigating}. 

Regarding the reconfigurable-based obfuscation techniques, one of the most formative works came from Microsoft and Iowa State University authors as describe in \cite{reconfigure_1}; the authors appropriately identified that reconfigurable logic could be leveraged as an obfuscation asset. Subsequently, a series of research studies emerged that utilize reconfigurable-based obfuscation schemes to protect digital designs. These techniques include a variety of reconfigurable elements, i.e., static random access memory (SRAM)-LUTs \cite{reconfigure_1, reconfigure_3, eASIC, kolhe2022silicon, lut1, Zain_TCAD, chowdhury2021enhancing}, NVM-LUTs \cite{attaran2018static, winograd2016hybrid, yang2018exploiting, lut2, lut3, kolhe2021securing, lut5}, and Others \cite{intro_epic, e-FPGA1, e-FPGA2, e-FPGA3, e-FPGA4, trap, mohan2021hardware, chen2021area, patnaik2018advancing, tomajoli2022alice, rangarajan2020opening, sathe2022investigating}. These reconfigurable techniques are very promising and demonstrate potential assurances against almost all hardware security threats. Overall, this paper is the \emph{first survey} to focus on \emph{reconfigurable-based obfuscation techniques} that combat security threats. Our intention is to present a detailed study of obfuscation trends, trade-offs, and recent attacks. This work provides a comprehensive overview of the reconfigurable-based obfuscation landscape, classifying the techniques based on three important factors: technology used, element type, and IP type. The efficiency of these techniques is evaluated and compared in terms of Power-Performance-Area (PPA) overheads. Finally, we discuss the benefits of reconfigurable-based obfuscation when compared to Logic Locking techniques and the challenges of evaluating obfuscation on hardware. 

The structure of this paper is given as follows: Section \ref{sec:background} classifies the reconfigurable-based obfuscation techniques and provides in-depth explanations. Section \ref{sec:Principles_com} elaborates on the comparison and analysis between numerous reconfigurable-based obfuscation techniques. Then, a comprehensive study of attacks and their evaluation of various obfuscation techniques is given in Section \ref{sec:securit_analysis}. In Section \ref{sec:discussion}, a rich discussion is provided as well as a comparison to Logic Locking. Future trends and challenges are also discussed in Section \ref{sec:discussion}. Finally, we conclude in Section \ref{sec:conclusions}.

%%%%%%%%%%%%%%%%%%%%%%%%%%%%%%%%%%%%%%%%%%%%%%%%%%%%%%%%%%%%%%%%%%%%%%%%%%%%%%%
%% SECTION - 2 (Background and classification of Reconfigurable-based Obfuscation)
%%%%%%%%%%%%%%%%%%%%%%%%%%%%%%%%%%%%%%%%%%%%%%%%%%%%%%%%%%%%%%%%%%%%%%%%%%%%%%%
\section{Background and Classification of Reconfigurable-based Obfuscation} \label{sec:background}
As previously mentioned, in order to have access to advanced technologies, design companies often outsource IC manufacturing to third-party foundries. Fabless design houses are concerned with protecting their designs against potential threats that could arise during manufacturing at an untrusted foundry. In this context, reconfigurable-based obfuscation techniques can safeguard designs by addressing nearly all of the threats outlined in Fig. \ref{fig:design_flow} \footnote{The effectiveness against side-channel attacks is not always present.}.

\subsection{Background} \label{subsec:background}

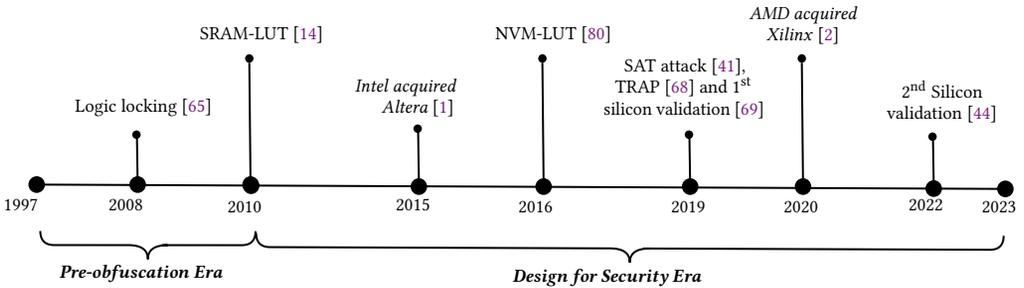
\begin {figure}[ht]
\begin{center}

\tikzset{every picture/.style={line width=0.75pt}} %set default line width to 0.75pt        

\begin{tikzpicture}[x=0.75pt,y=0.75pt,yscale=-1,xscale=1]

%Shape: Brace [id:dp1358595175394337] 
\draw   (37,156.33) .. controls (37,161) and (39.33,163.33) .. (44,163.34) -- (81.33,163.36) .. controls (88,163.37) and (91.33,165.7) .. (91.32,170.37) .. controls (91.33,165.7) and (94.66,163.37) .. (101.33,163.38)(98.33,163.38) -- (138.66,163.4) .. controls (143.33,163.41) and (145.66,161.08) .. (145.67,156.41) ;
%Shape: Brace [id:dp2825408005771215] 
\draw   (146,157.33) .. controls (145.98,162) and (148.3,164.34) .. (152.97,164.36) -- (324.71,164.99) .. controls (331.38,165.02) and (334.7,167.36) .. (334.68,172.02) .. controls (334.7,167.36) and (338.04,165.04) .. (344.71,165.06)(341.71,165.05) -- (516.44,165.69) .. controls (521.11,165.71) and (523.45,163.39) .. (523.47,158.72) ;
%Shape: Circle [id:dp663747188683844] 
\draw  [draw opacity=0][fill={rgb, 255:red, 0; green, 0; blue, 0 }  ,fill opacity=1 ] (31,133.17) .. controls (31,130.87) and (32.87,129) .. (35.17,129) .. controls (37.47,129) and (39.33,130.87) .. (39.33,133.17) .. controls (39.33,135.47) and (37.47,137.33) .. (35.17,137.33) .. controls (32.87,137.33) and (31,135.47) .. (31,133.17) -- cycle ;
%Straight Lines [id:da07508501048180038] 
\draw    (35.17,133.17) -- (524.33,135.33) ;
%Shape: Circle [id:dp6550768091321935] 
\draw  [draw opacity=0][fill={rgb, 255:red, 0; green, 0; blue, 0 }  ,fill opacity=1 ] (520.17,135.33) .. controls (520.17,133.03) and (522.03,131.17) .. (524.33,131.17) .. controls (526.63,131.17) and (528.5,133.03) .. (528.5,135.33) .. controls (528.5,137.63) and (526.63,139.5) .. (524.33,139.5) .. controls (522.03,139.5) and (520.17,137.63) .. (520.17,135.33) -- cycle ;
%Shape: Circle [id:dp9764061292294863] 
\draw  [draw opacity=0][fill={rgb, 255:red, 0; green, 0; blue, 0 }  ,fill opacity=1 ] (82,133.17) .. controls (82,130.87) and (83.87,129) .. (86.17,129) .. controls (88.47,129) and (90.33,130.87) .. (90.33,133.17) .. controls (90.33,135.47) and (88.47,137.33) .. (86.17,137.33) .. controls (83.87,137.33) and (82,135.47) .. (82,133.17) -- cycle ;
%Shape: Circle [id:dp005861857366561196] 
\draw  [draw opacity=0][fill={rgb, 255:red, 0; green, 0; blue, 0 }  ,fill opacity=1 ] (484,135.17) .. controls (484,132.87) and (485.87,131) .. (488.17,131) .. controls (490.47,131) and (492.33,132.87) .. (492.33,135.17) .. controls (492.33,137.47) and (490.47,139.33) .. (488.17,139.33) .. controls (485.87,139.33) and (484,137.47) .. (484,135.17) -- cycle ;
%Shape: Circle [id:dp9347886463633539] 
\draw  [draw opacity=0][fill={rgb, 255:red, 0; green, 0; blue, 0 }  ,fill opacity=1 ] (418,134.17) .. controls (418,131.87) and (419.87,130) .. (422.17,130) .. controls (424.47,130) and (426.33,131.87) .. (426.33,134.17) .. controls (426.33,136.47) and (424.47,138.33) .. (422.17,138.33) .. controls (419.87,138.33) and (418,136.47) .. (418,134.17) -- cycle ;
%Shape: Circle [id:dp9567017744635888] 
\draw  [draw opacity=0][fill={rgb, 255:red, 0; green, 0; blue, 0 }  ,fill opacity=1 ] (361,134.17) .. controls (361,131.87) and (362.87,130) .. (365.17,130) .. controls (367.47,130) and (369.33,131.87) .. (369.33,134.17) .. controls (369.33,136.47) and (367.47,138.33) .. (365.17,138.33) .. controls (362.87,138.33) and (361,136.47) .. (361,134.17) -- cycle ;
%Shape: Circle [id:dp9984310152454952] 
\draw  [draw opacity=0][fill={rgb, 255:red, 0; green, 0; blue, 0 }  ,fill opacity=1 ] (287,134.17) .. controls (287,131.87) and (288.87,130) .. (291.17,130) .. controls (293.47,130) and (295.33,131.87) .. (295.33,134.17) .. controls (295.33,136.47) and (293.47,138.33) .. (291.17,138.33) .. controls (288.87,138.33) and (287,136.47) .. (287,134.17) -- cycle ;
%Shape: Circle [id:dp15940755637029969] 
\draw  [draw opacity=0][fill={rgb, 255:red, 0; green, 0; blue, 0 }  ,fill opacity=1 ] (224,134.17) .. controls (224,131.87) and (225.87,130) .. (228.17,130) .. controls (230.47,130) and (232.33,131.87) .. (232.33,134.17) .. controls (232.33,136.47) and (230.47,138.33) .. (228.17,138.33) .. controls (225.87,138.33) and (224,136.47) .. (224,134.17) -- cycle ;
%Shape: Circle [id:dp18546758296374555] 
\draw  [draw opacity=0][fill={rgb, 255:red, 0; green, 0; blue, 0 }  ,fill opacity=1 ] (139.3,133.73) .. controls (139.3,131.43) and (141.17,129.57) .. (143.47,129.57) .. controls (145.77,129.57) and (147.63,131.43) .. (147.63,133.73) .. controls (147.63,136.03) and (145.77,137.9) .. (143.47,137.9) .. controls (141.17,137.9) and (139.3,136.03) .. (139.3,133.73) -- cycle ;
%Straight Lines [id:da2546220345881749] 
\draw    (487.78,109.89) -- (488.17,135.17) ;
%Shape: Circle [id:dp04214695176276306] 
\draw  [draw opacity=0][fill={rgb, 255:red, 0; green, 0; blue, 0 }  ,fill opacity=1 ] (485.43,109.28) .. controls (485.43,108.11) and (486.38,107.17) .. (487.55,107.17) .. controls (488.72,107.17) and (489.67,108.11) .. (489.67,109.28) .. controls (489.67,110.45) and (488.72,111.4) .. (487.55,111.4) .. controls (486.38,111.4) and (485.43,110.45) .. (485.43,109.28) -- cycle ;
%Straight Lines [id:da9278123704120036] 
\draw    (364.78,108.89) -- (365.17,134.17) ;
%Shape: Circle [id:dp5559523179989971] 
\draw  [draw opacity=0][fill={rgb, 255:red, 0; green, 0; blue, 0 }  ,fill opacity=1 ] (362.43,108.28) .. controls (362.43,107.11) and (363.38,106.17) .. (364.55,106.17) .. controls (365.72,106.17) and (366.67,107.11) .. (366.67,108.28) .. controls (366.67,109.45) and (365.72,110.4) .. (364.55,110.4) .. controls (363.38,110.4) and (362.43,109.45) .. (362.43,108.28) -- cycle ;
%Straight Lines [id:da1767348539442215] 
\draw    (227.78,105.89) -- (228.01,120.85) -- (228.17,131.17) ;
%Shape: Circle [id:dp10637258346834333] 
\draw  [draw opacity=0][fill={rgb, 255:red, 0; green, 0; blue, 0 }  ,fill opacity=1 ] (225.43,105.28) .. controls (225.43,104.11) and (226.38,103.17) .. (227.55,103.17) .. controls (228.72,103.17) and (229.67,104.11) .. (229.67,105.28) .. controls (229.67,106.45) and (228.72,107.4) .. (227.55,107.4) .. controls (226.38,107.4) and (225.43,106.45) .. (225.43,105.28) -- cycle ;
%Straight Lines [id:da78238756950667] 
\draw    (85.78,108.89) -- (86.17,134.17) ;
%Shape: Circle [id:dp6524479586857612] 
\draw  [draw opacity=0][fill={rgb, 255:red, 0; green, 0; blue, 0 }  ,fill opacity=1 ] (83.43,108.28) .. controls (83.43,107.11) and (84.38,106.17) .. (85.55,106.17) .. controls (86.72,106.17) and (87.67,107.11) .. (87.67,108.28) .. controls (87.67,109.45) and (86.72,110.4) .. (85.55,110.4) .. controls (84.38,110.4) and (83.43,109.45) .. (83.43,108.28) -- cycle ;
%Shape: Circle [id:dp6335806856745887] 
\draw  [draw opacity=0][fill={rgb, 255:red, 0; green, 0; blue, 0 }  ,fill opacity=1 ] (139,133.17) .. controls (139,130.87) and (140.87,129) .. (143.17,129) .. controls (145.47,129) and (147.33,130.87) .. (147.33,133.17) .. controls (147.33,135.47) and (145.47,137.33) .. (143.17,137.33) .. controls (140.87,137.33) and (139,135.47) .. (139,133.17) -- cycle ;
%Straight Lines [id:da45341811585383685] 
\draw    (142.55,72.4) -- (143.17,134.17) ;
%Shape: Circle [id:dp5078889523356624] 
\draw  [draw opacity=0][fill={rgb, 255:red, 0; green, 0; blue, 0 }  ,fill opacity=1 ] (140.43,70.28) .. controls (140.43,69.11) and (141.38,68.17) .. (142.55,68.17) .. controls (143.72,68.17) and (144.67,69.11) .. (144.67,70.28) .. controls (144.67,71.45) and (143.72,72.4) .. (142.55,72.4) .. controls (141.38,72.4) and (140.43,71.45) .. (140.43,70.28) -- cycle ;
%Straight Lines [id:da8923203198512502] 
\draw    (290.55,72.4) -- (291.17,134.17) ;
%Shape: Circle [id:dp36459071057546444] 
\draw  [draw opacity=0][fill={rgb, 255:red, 0; green, 0; blue, 0 }  ,fill opacity=1 ] (288.43,70.28) .. controls (288.43,69.11) and (289.38,68.17) .. (290.55,68.17) .. controls (291.72,68.17) and (292.67,69.11) .. (292.67,70.28) .. controls (292.67,71.45) and (291.72,72.4) .. (290.55,72.4) .. controls (289.38,72.4) and (288.43,71.45) .. (288.43,70.28) -- cycle ;
%Straight Lines [id:da870283144694965] 
\draw    (421.55,72.4) -- (422.17,134.17) ;
%Shape: Circle [id:dp6392360238139725] 
\draw  [draw opacity=0][fill={rgb, 255:red, 0; green, 0; blue, 0 }  ,fill opacity=1 ] (419.43,70.28) .. controls (419.43,69.11) and (420.38,68.17) .. (421.55,68.17) .. controls (422.72,68.17) and (423.67,69.11) .. (423.67,70.28) .. controls (423.67,71.45) and (422.72,72.4) .. (421.55,72.4) .. controls (420.38,72.4) and (419.43,71.45) .. (419.43,70.28) -- cycle ;

% Text Node
\draw (17,138.53) node [anchor=north west][inner sep=0.75pt]  [font=\scriptsize] [align=left] {1997};
% Text Node
\draw (53,88.53) node [anchor=north west][inner sep=0.75pt]  [font=\scriptsize] [align=left] {\begin{minipage}[lt]{52pt}\setlength\topsep{0pt}
\begin{center}
Logic locking \cite{intro_epic}
\end{center}

\end{minipage}};
% Text Node
\draw (114,52.53) node [anchor=north west][inner sep=0.75pt]  [font=\scriptsize] [align=left] {\begin{minipage}[lt]{50pt}\setlength\topsep{0pt}
\begin{center}
SRAM-LUT \cite{reconfigure_1}
\end{center}

\end{minipage}};
% Text Node
\draw (386,42.53) node [anchor=north west][inner sep=0.75pt]  [font=\scriptsize] [align=left] {\begin{minipage}[lt]{51.94pt}\setlength\topsep{0pt}
\begin{center}
\textit{AMD acquired \\Xilinx} \cite{amdacquiredxilinx}
\end{center}

\end{minipage}};
% Text Node
\draw (320,68.53) node [anchor=north west][inner sep=0.75pt]  [font=\scriptsize] [align=left] {\begin{minipage}[lt]{60.88pt}\setlength\topsep{0pt}
\begin{center}
SAT attack \cite{lut3}, TRAP \cite{e-FPGA3} and 1\textsuperscript{st} silicon validation \cite{trap}
\end{center}

\end{minipage}};
% Text Node
\draw (457,79.53) node [anchor=north west][inner sep=0.75pt]  [font=\scriptsize] [align=left] {\begin{minipage}[lt]{50.89pt}\setlength\topsep{0pt}
\begin{center}
2\textsuperscript{nd} Silicon \\validation \cite{kolhe2022silicon}
\end{center}

\end{minipage}};
% Text Node
\draw (194.47,78.73) node [anchor=north west][inner sep=0.75pt]  [font=\scriptsize] [align=left] {\begin{minipage}[lt]{47.57pt}\setlength\topsep{0pt}
\textit{Intel acquired} 
\begin{center}
\textit{Altera} \cite{intelacquiredaltera}
\end{center}

\end{minipage}};
% Text Node
\draw (261,52.53) node [anchor=north west][inner sep=0.75pt]  [font=\scriptsize] [align=left] {\begin{minipage}[lt]{50.45pt}\setlength\topsep{0pt}
\begin{center}
NVM-LUT \cite{winograd2016hybrid}
\end{center}

\end{minipage}};
% Text Node
\draw (45,171.33) node [anchor=north west][inner sep=0.75pt]  [font=\scriptsize] [align=left] {\textbf{\textit{Pre-obfuscation Era}}};
% Text Node
\draw (274,173.33) node [anchor=north west][inner sep=0.75pt]  [font=\scriptsize] [align=left] {\textbf{\textit{Design for Security Era}}};
% Text Node
\draw (70,138.53) node [anchor=north west][inner sep=0.75pt]  [font=\scriptsize] [align=left] {2008};
% Text Node
\draw (130,139.53) node [anchor=north west][inner sep=0.75pt]  [font=\scriptsize] [align=left] {2010};
% Text Node
\draw (215,138.53) node [anchor=north west][inner sep=0.75pt]  [font=\scriptsize] [align=left] {2015};
% Text Node
\draw (277,139.53) node [anchor=north west][inner sep=0.75pt]  [font=\scriptsize] [align=left] {2016};
% Text Node
\draw (354,139.53) node [anchor=north west][inner sep=0.75pt]  [font=\scriptsize] [align=left] {2019};
% Text Node
\draw (411,139.53) node[anchor=north west][inner sep=0.75pt]  [font=\scriptsize] [align=left] {2020};
% Text Node
\draw (474,138.53) node [anchor=north west][inner sep=0.75pt]  [font=\scriptsize] [align=left] {2022};
% Text Node
\draw (511,139.53) node [anchor=north west][inner sep=0.75pt]  [font=\scriptsize] [align=left] {2023};
\end{tikzpicture}
\caption{Reconfigurable Logic: Navigating the Shift from Pre-Obfuscation to Security Era}
\label{fig:timeline}
\end{center}
\end{figure}

Over the last few decades, reconfigurable devices such as FPGAs and FPGA-based SoCs have been widely used as stand-alone solutions. It was only recently that the ability to reconfigure a design has been considered a form of obfuscation. In Fig. \ref{fig:timeline}, we frame the evolution and usage of reconfigurable devices into two phases: the \textbf{pre-obfuscation era} and the \textbf{design for security era}. In 1984, Xilinx introduced the first FPGA, called the XC2064 \cite{history_fpga}. This FPGA had 64 logic cells and was programmed using a hardware description language (HDL) called ABEL. In the late 1980s and early 1990s, FPGAs became more popular as their capacity increased and their price decreased \cite{history_fpga_detail}. Xilinx and Altera were the two leading FPGA manufacturers at that time. 

\begin{figure*}[tb!]
    \centerline{\includegraphics[width=0.4\linewidth]{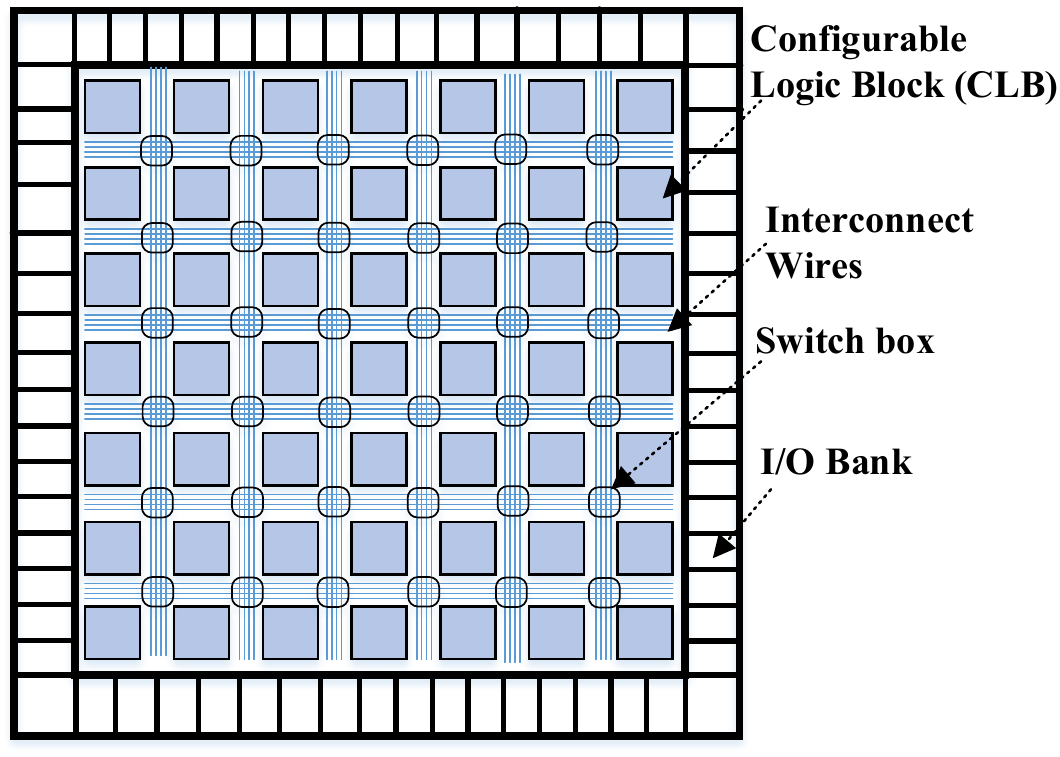}}
    \caption{The traditional island-style architecture of FPGA (adapted from \cite{only_fpga}).}
    \label{fig:fpga_architecture}
\end{figure*}

The traditional island-style architecture of an FPGA is illustrated in Fig. \ref{fig:fpga_architecture}. The architecture of an FPGA consists of several basic building blocks such as interconnect wires, configurable logic block (CLB), switch matrix, and I/O bank. A CLB consists of a small number of logic gates and can be configured to implement a specific logic function. %The CLB can also be programmed to implement more complex functions by connecting multiple CLBs together through a programmable interconnect network.
The input/output blocks (IOBs) provide the interface between the FPGA and the external world. %They are used to input and output data to and from the FPGA, and can be configured to support a wide range of interfaces, such as USB, Ethernet, and PCI. The programmable interconnect network is another key feature of an FPGA architecture. 
The interconnect network is used to connect the various building blocks together, and can be programmed to implement different routing configurations depending on the specific application. A CLB is typically composed of a LUT, a flip-flop, and a multiplexer. The LUT is used to implement combinational logic functions, while the flip-flop is used to store the state of a sequential logic circuit. The multiplexer is used to select between different inputs to the logic element. However, modern FPGAs are characterized by more complex and non-uniform device grids, with I/Os often placed in columns instead of around the perimeter \cite{routing_effect}. It has been observed that the size of LUTs in FPGAs has evolved over time. Accordingly, the Virtex 4 family of FPGAs had 4-input LUTs, whereas the Virtex 5 and Virtex 6 families had 5-input and 6-input LUTs, respectively \cite{virtex5, virtex6}. It is even possible to find some manufacturers offering FPGAs with even larger 8-input LUTs \cite{lut8}. With the evolution of LUT sizes, FPGAs are able to support a wide range of increasingly complex and sophisticated digital logic designs. Except for minor terminology differences among different vendors, the architecture depicted in Fig.~\ref{fig:fpga_architecture} is representative of an FPGA. %Overall, the architecture of an FPGA is designed to be highly flexible and reconfigurable, allowing it to be programmed to implement a wide range of digital functions. Missing

%%%%%%%%%%%%%%%%%%%%%%%%%%%%%%%%%%%%%%%%%%%%%%%%%%%%%%%%%%%%%%%%%%%%%%%%%%%%%%%
%% FIGURE 
%%%%%%%%%%%%%%%%%%%%%%%%%%%%%%%%%%%%%%%%%%%%%%%%%%%%%%%%%%%%%%%%%%%%%%%%%%%%%%%

In the early 2000s, new application domains emerged and the underlying process technology changed while the architecture of FPGAs continued to evolve. The ability to partially program and dynamically reconfigure FPGA architectures enables digital circuit design and implementation to be more flexible. Modifying only a small portion of the fabric allows for the design to continue to operate while some small parts are programmed. In practice, the same board can be used to switch from one design to another design, allowing for greater design flexibility \cite{partial_reconfigure}. In FPGA-based system on chips (SoCs), partial and dynamic reconfiguration \cite{dynamic_reconfigure} are possible for the FPGA fabric as well as for periphery IPs.

%Partial reconfiguration is typically accomplished by dividing the FPGA into multiple regions, with each region containing a specific subset of the FPGA fabric that can be modified independently.
%On the other hand, in dynamic reconfiguration, hardware-based techniques such as partial reconfiguration are typically used, where only a portion of the FPGA is changed at a time . Partial reconfiguration allows for greater design flexibility and faster design iterations, while dynamic reconfiguration allows faster and more adaptable FPGA designs by modifying the fabric on-the-fly. %This can be particularly useful in embedded systems and high-speed data communication applications to enhance real-time performance. Overall, partial and dynamic reconfiguration are important features of modern FPGA architectures that enable greater design flexibility and efficiency. These features have helped to make FPGAs a popular choice for a wide range of applications, from high-performance computing to embedded systems and beyond.

Moving forward with the evolution of FPGA architectures, modern FPGAs are comprised of a range of modules including memory, DSP, PLLs, clocking, networking, and more \cite{xilinx_fpga, altera_fpga}. Looking ahead, these blocks are expected to continue evolving and expanding in capability. Examples of these architectures include large blocks such as hardware accelerators \cite{xilinx_accelator}. A hybrid architecture emerged in 2003 when Xilinx embedded its FPGA technology into IBM's Application-Specific Integrated Circuits (ASICs), enabling designers to add programmable logic into their designs without developing a separate board for the FPGA \cite{edn}. At the same time, FPGA-based SoCs targeting digital signal processing applications extended their computational power through reconfigurable solutions often involving a matrix of computational elements with programmable interconnections \cite {singh2000morphosys, zhang20001, becker2001parallel, borgatti2003reconfigurable}. In practice, FPGA-based SoC are complex devices that employ much more than hard and soft embedded processors \cite{xilinx_socs}. In a very recent trend, FPGA-based SoCs are being shipped with a Network-on-Chip (NoC) subsystem \cite{fpga_noc} to interconnect all of its modules.

Over the next two decades, the boundaries between ASIC and FPGA design became less clear. The embedded-Field Programmable Gate Arrays (eFPGAs) emerged as a small FPGA module that can be integrated into ASIC. The eFPGA IP can be licensed for use in a similar manner to any other IP. For each application, eFPGA IP designers can specify exactly how many logic units, digital signal processing (DSP) and machine learning processing units (MLP) they provide. By eliminating unnecessary FPGA features, this results in more flexibility, lower costs, and smaller eFPGA IP area. Furthermore, if a custom or special FPGA architecture is needed, it can also be implemented as an eFPGA for greater reconfigurability. Several IP providers are offering eFPGA blocks of various granularities and architectures. 

Concerning the security side, a landmark development in the field of obfuscation is the Logic Locking concept \cite {intro_epic} that came to light in 2008 (see timeline in Fig. \ref{fig:timeline}). Logic locking is a technique used at design time to safeguard ICs against supply chain threats and it involves the insertion of additional logic into a circuit and securing it with a secret key. Key inputs, driven by an on-chip tamper-proof memory, are added to the locked circuit along with the original inputs. The additional logic may consist of combinational logic like MUX, AND, OR, and XOR gates \cite{logic_3}. The design functions correctly and produces the correct output only when the correct key value is applied. Otherwise, its output differs from that of the original design. 
% Divided into paragraphs

\begin{figure*} [ht]
     \centering
     \begin{subfigure}[b]{0.31\textwidth}
         \centering
         \includegraphics[width=\textwidth]{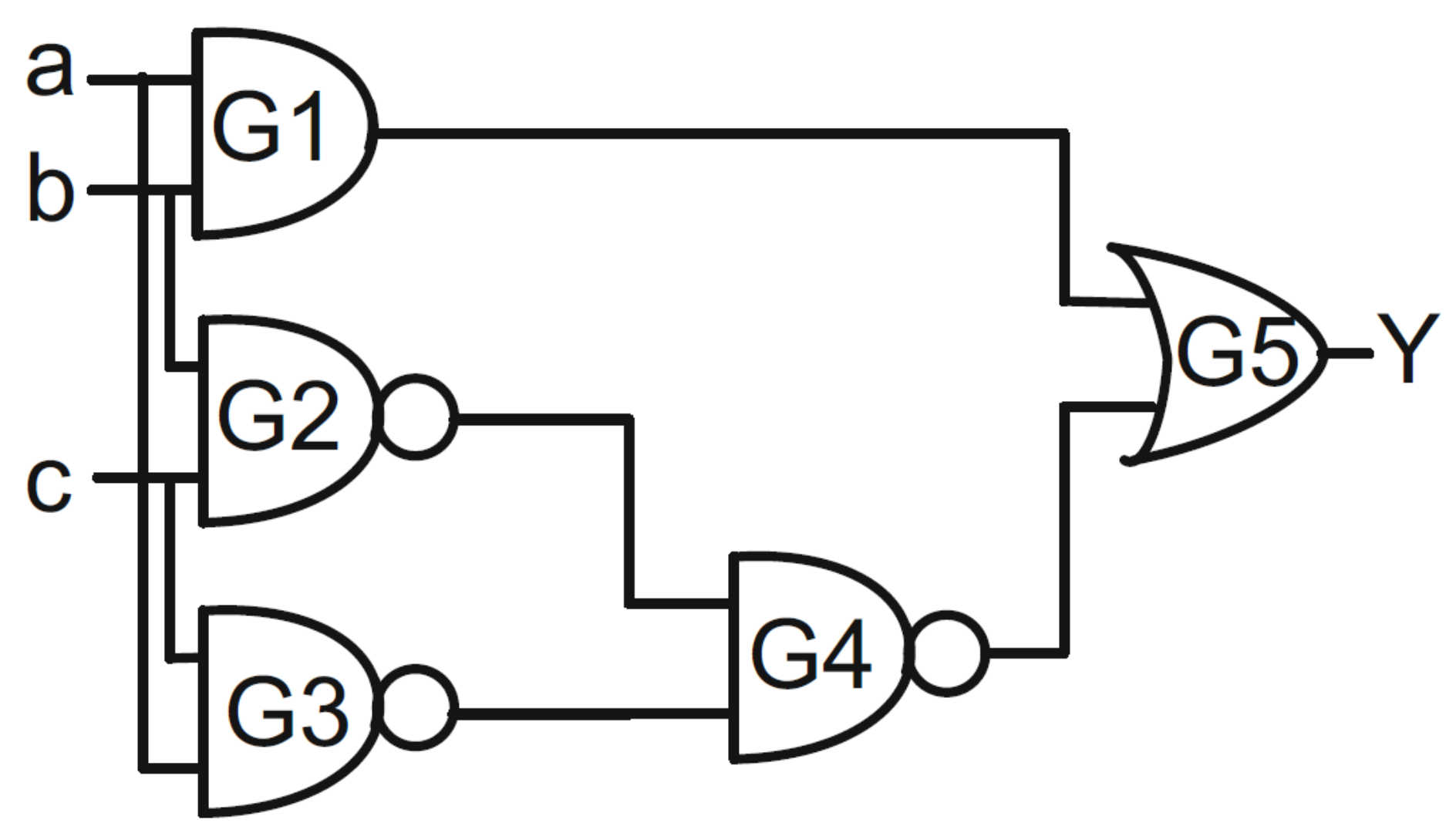}
         \caption{An example circuit that consists of three inputs.}
         \label{fig:LL_a}
     \end{subfigure}
     %\hfill
     \hspace{1em}
     \begin{subfigure}[b]{0.38\textwidth}
         \centering
         \includegraphics[width=\textwidth]{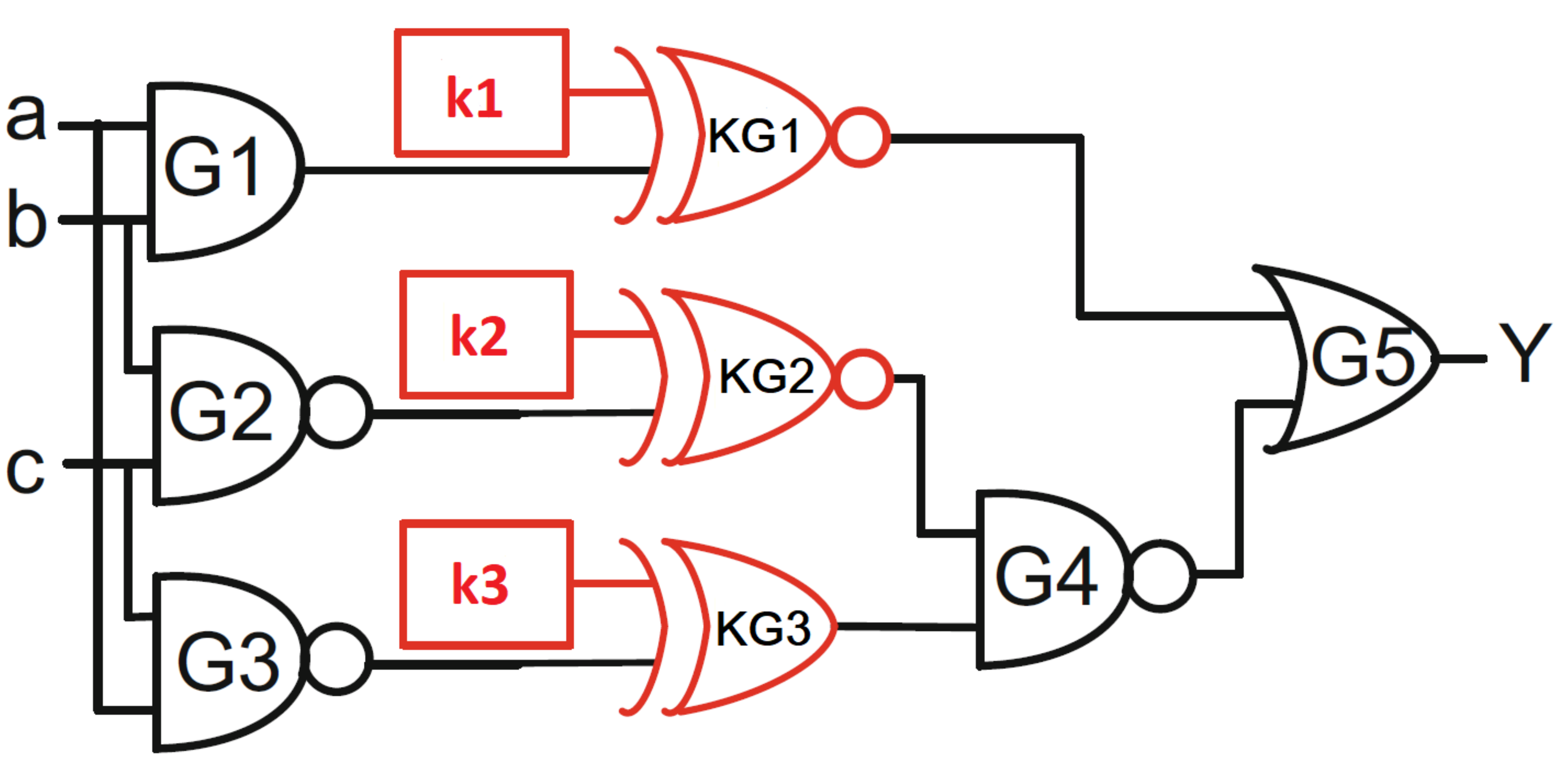}
         \caption{Three XOR/XNOR gates should use the key value of 110.}
         \label{fig:LL_b}
     \end{subfigure}
     \hfill
        \caption{Logic locking using XOR/XNOR gates \cite{logic_3}.}
        \label{fig:logic_locking}
\end{figure*}

In Fig. \ref{fig:LL_a}, an example of a circuit that consists of three inputs and one output is depicted. Fig. \ref{fig:LL_b} shows the locked version of the circuit, which includes three additional XOR/XNOR key gates. Each key gate has one input driven by a wire of the original design, while the other input, called the key input, is driven by a key bit stored in the tamper-proof memory. In the locked circuit of Fig. \ref{fig:LL_b}, when the correct key value 110 is loaded into memory, all key gates behave as buffers and produce the correct output for any input pattern. However, applying an incorrect key value, such as 010, causes certain key gates to behave as inverters, leading to an error injection into the circuit. For instance, the key gate K1 acts as an inverter in the case of input pattern 000 and key value 010, producing an incorrect output Y = 1.

%%%%%%%%%%%%%%%%%%%%%%%%%%%%%%%%%%%%%%%%%%%%%%%%%%%%%%%%%%%%%%%%%%%%%%%%%%%%%%%
%% FIGURE 
%%%%%%%%%%%%%%%%%%%%%%%%%%%%%%%%%%%%%%%%%%%%%%%%%%%%%%%%%%%%%%%%%%%%%%%%%%%%%%%

This initial period in the evolution of reconfigurable devices can be considered as \textbf{the pre-obfuscation era}. Then, the first reconfigurable-based obfuscation technique was proposed in 2010. This marks the beginning of the \textbf{the design for security era} which continues to this day. Several techniques have been proposed since.

Recent years were also marked by significant acquisitions in the semiconductors market. One of such remarkable events is Intel's acquisition of Altera, a leading provider of FPGA technology, in 2015 \cite{intel_acquired}. By acquiring Altera, Intel gained access to the company's industry-leading FPGA technology, which is widely used in data centers, networking, and embedded systems applications. In 2020, AMD's acquisition of Xilinx was a strategic move to expand its market reach and strengthen its position in the high-performance computing (HPC) and data center markets. Xilinx FPGAs and adaptive SoC solutions complement AMD's portfolio of central processing units (CPUs), graphics processing units (GPUs), and other accelerator technologies.

%From 2010 to 2015, a few reconfigurable-based techniques were proposed, as reported in \cite{reconfigure_3}.

In 2016, Menta introduced the first commercially available eFPGA IP, which allows designers to integrate eFPGA IP into their own ASICs. This approach enables the integration of a reprogrammable logic fabric into a wide range of ASICs. Another reconfigurable-based obfuscation was introduced in the same year, which utilized LUTs to obfuscate the design. Several reconfigurable-based obfuscation techniques which exploit LUTs to hide circuits were proposed from 2018 to 2019. The year 2019 marked a significant breakthrough, with the introduction of another reconfigurable-based obfuscation technique where the transistors are programmed to recover the functionality of the circuit. In the same year, the first Boolean satisfiability problem (SAT) attack on reconfigurable-based obfuscation was also introduced, which marked a notable development in the field. 

\begin{figure*}[ht]
    \centerline{\includegraphics[width=0.80\linewidth]{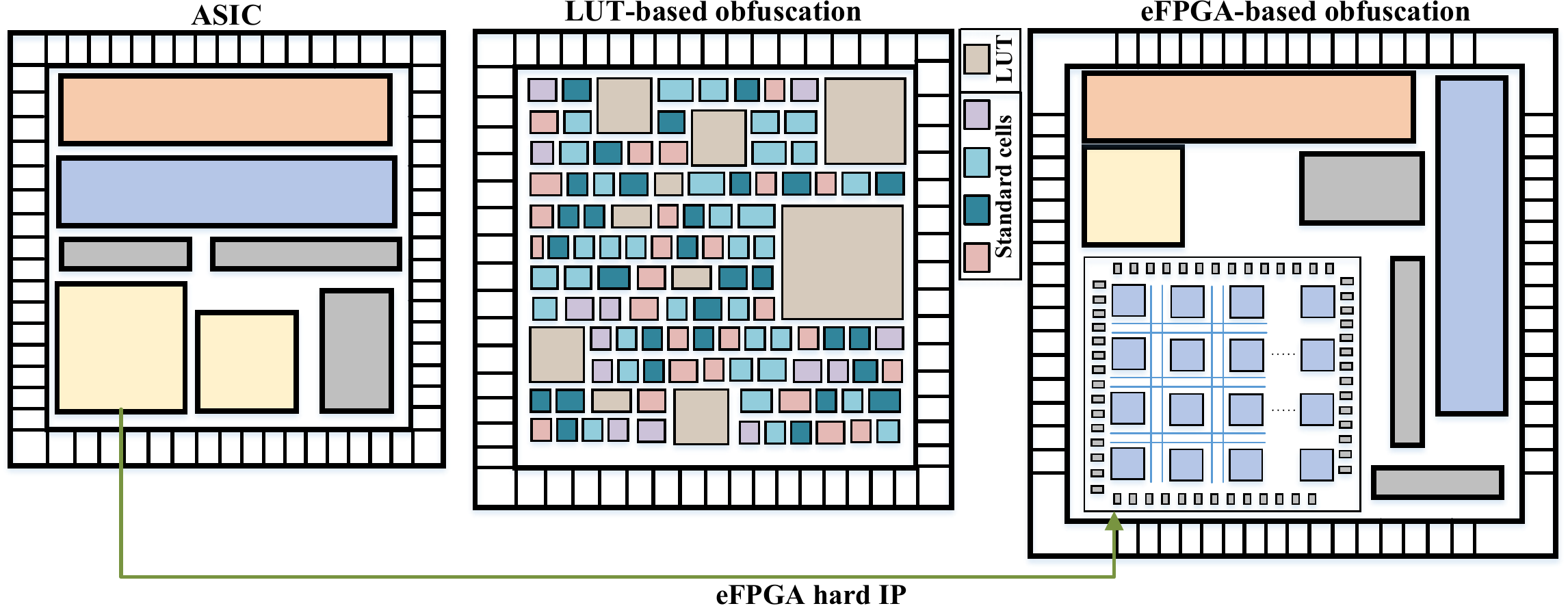}}
    \caption{Interpretation of eFPGA-based obfuscation technique \cite{tomajoli2022alice}. Notice the increase in die area and the modified floorplan. }
    \label{fig:FPGA_redaction}
\end{figure*}

In that same year, a novel reconfigurable-based obfuscation technique was proposed that utilized eFPGAs for obfuscation purposes \cite{e-FPGA2}. Researchers referred to this as``eFPGA redaction'' which is another term used to describe reconfigurable-based obfuscation techniques that utilize eFPGAs as an obfuscation asset. An example of eFPGA redaction is illustrated in the right panel of Fig. \ref{fig:FPGA_redaction} which leverages an eFPGA macro to obfuscate the circuit. This level of obfuscation is inserted during the physical synthesis of a design, therefore this type of technique demands certain skills in physical implementation that are far more complicated than the insertion of XOR/XNOR in a netlist (as illustrated in Fig. \ref{fig:logic_locking}). In summary, the final layout will look like the one illustrated in the right panel of Fig. \ref{fig:FPGA_redaction}. Recalling again, the generated layout must be shared with the untrusted foundry for manufacturing. However, the bitstream for the eFPGA IP will not be shared with the untrusted foundry. The example highlighted in Fig. \ref{fig:FPGA_redaction} demonstrates how a single module is turned into a reconfigurable part to offer obfuscation. 

Between 2010 to 2020, there were multiple reconfigurable-based obfuscation techniques that were offering obfuscation with a high level of security. However, none of them have demonstrated silicon validation. In 2020, the authors of \cite{kolhe2022silicon} demonstrated what is likely the first proof of concept in silicon. It is noteworthy that the traditional FPGA uses SRAM bit-cells to reconfigure logic. In this work, however, authors utilized daisy-chained flip flops to keep the bitstream. %In the next section, we will provide further information on how the reconfigurable part can either be integrated as a single module or distributed throughout the design. 

%%%%%%%%%%%%%%%%%%%%%%%%%%%%%%%%%%%%%%%%%%%%%%%%%%%%%%%%%%%%%%%%%%%%%%%%%%%%%%%
%% FIGURE 
%%%%%%%%%%%%%%%%%%%%%%%%%%%%%%%%%%%%%%%%%%%%%%%%%%%%%%%%%%%%%%%%%%%%%%%%%%%%%%%

%Today, embedded FPGAs are used in ASICs and number continue to grow as shown in the Fig. \ref{}a variety of industries and applications, including wireless communication, networking, industrial automation, and artificial intelligence. These changes and additions to the FPGA architecture are a result of the increasing demand for more advanced and versatile FPGAs capable of meeting the requirements of modern and emerging application domains.\cite {baumgarte2001pact}. Even if designs were equipped with reconfigurable assets, these were present only as a design aid and not as a security asset. %In the following paragraphs, we compare and discuss the effectiveness of reconfigurable devices for hardware security. 

\begin {figure}[h!]
\centerline{\includegraphics[width=0.55\linewidth]{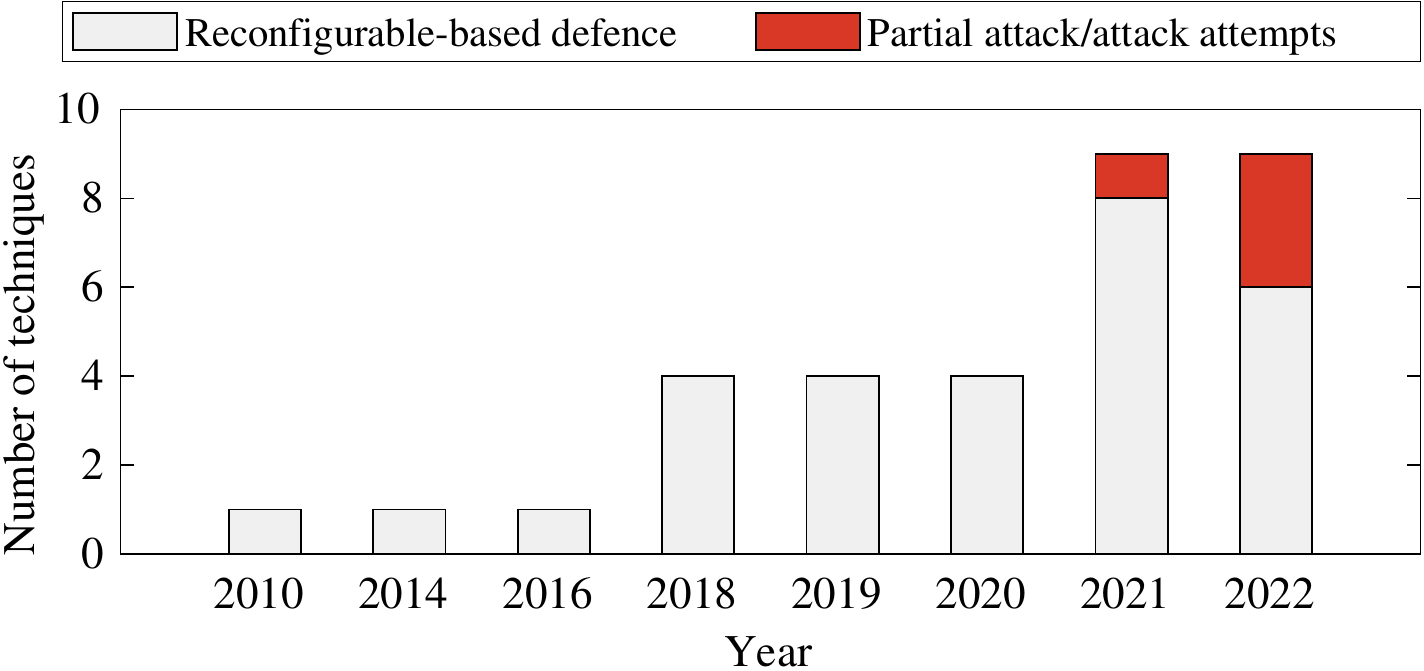}}
\caption{Publication trend and attacks for  reconfigurable-based obfuscation techniques.}
\label{fig:survey_trend}
\end{figure}

In order to summarize the advances in reconfigurable-based obfuscation, we present the publication trend of  techniques and attacks in Fig. \ref{fig:survey_trend}. It is clear from the trend that a great number of reconfigurable-based obfuscation techniques have been proposed to protect the IP against various hardware security attacks \cite{attaran2018static, reconfigure_1, reconfigure_3, lut1, patnaik2018advancing, winograd2016hybrid, yang2018exploiting}. The trend of defense techniques is continuously growing, and the research community has recently displayed significant interest in the attacks. Therefore, there are initial attempts at breaking obfuscation schemes based on reconfigurable devices. In most cases, even when using state-of-the-art attacks, adversaries appear to only be able to analyze the behavior of obfuscated circuits without much success. We elaborate on the details of these attacks in Section \ref{sec:securit_analysis}. Moreover, the substantial amount of publications has enabled us to classify the reconfigurable-based obfuscation techniques, which we further interpret in Section \ref{subsec:classifications}. %However, LUT-based obfuscation has received the most attention due to its ability to realize many logic elements with non-recurring engineering costs. For any SRAM LUT-based obfuscation \cite{reconfigure_1, reconfigure_3}, the selected internal gates from the design are mapped to the LUTs. For instance, to obfuscate a 2-input AND gate with LUT, one can replace the AND gate in the IP with the LUT whose configuration bits are set to “0001.” Therefore, LUT-based obfuscation produces a netlist that is a combination of ASIC and programmable FPGA. Here, the configuration bits or the keys that indicate the logical function of the LUT is stored in a tamper-proof non-volatile memory. Without prior knowledge of the key bits stored in the non-volatile memory, the attacker does not have access to the IP’s intended functionality and thus refrains the attacker from reverse engineering the IP.  

%%%%%%%%%%%%%%%%%%%%%%%%%%%%%%%%%%%%%%%%%%%%%%%%%%%%%%%%%%%%%%%%%%%%%%%%%%%%%%%
%% FIGURE 
%%%%%%%%%%%%%%%%%%%%%%%%%%%%%%%%%%%%%%%%%%%%%%%%%%%%%%%%%%%%%%%%%%%%%%%%%%%%%%%

\subsection{Classification of reconfigurable-based obfuscation techniques}
\label{subsec:classifications}
\begin {figure}[htb]
\begin{center}
\tikzset{every picture/.style={line width=0.75pt}} %set default line width to 0.75pt
\resizebox{0.65\textwidth}{!}{%
\begin{tikzpicture}[x=0.75pt,y=0.72pt,yscale=-1,xscale=1]
%uncomment if require: \path (0,676); %set diagram left start at 0, and has height of 676
%Shape: Rectangle [id:dp5275842828045727] 
% Technology used box
\draw   (87,102.05) -- (200,102.05) -- (200,143) -- (87,143) -- cycle ;
%Shape: Rectangle [id:dp6919324169632739] 
% SRAm LUT
\draw   (111.85,163.86) -- (236,163.86) -- (236,251.86) -- (111.85,251.86) -- cycle ;
%Shape: Rectangle [id:dp3780288976865338] 
% Reconfigurable-based obfuscation
\draw   (238.99,7) -- (415.93,7) -- (415.93,54.14) -- (238.99,54.14) -- cycle ;
%Straight Lines [id:da9713910912933494] 
% Arrow
\draw    (327.84,54.14) -- (327.84,100.05) ;
\draw [shift={(327.84,102.05)}, rotate = 270] [color={rgb, 255:red, 0; green, 0; blue, 0 }  ][line width=0.75]    (10.93,-3.29) .. controls (6.95,-1.4) and (3.31,-0.3) .. (0,0) .. controls (3.31,0.3) and (6.95,1.4) .. (10.93,3.29)   ;
%Straight Lines [id:da9913711712057223] 
% Lines
\draw    (157.02,78.86) -- (490.68,78.09) ;
%Straight Lines [id:da1872984787150005] 
\draw    (157.02,78.86) -- (157.02,100.43) ;
\draw [shift={(157.02,102.43)}, rotate = 270] [color={rgb, 255:red, 0; green, 0; blue, 0 }  ][line width=0.75]    (10.93,-3.29) .. controls (6.95,-1.4) and (3.31,-0.3) .. (0,0) .. controls (3.31,0.3) and (6.95,1.4) .. (10.93,3.29)   ;
%Straight Lines [id:da4614117599991654] 
\draw    (490.68,78.09) -- (490.68,99.27) ;
\draw [shift={(490.68,101.27)}, rotate = 270] [color={rgb, 255:red, 0; green, 0; blue, 0 }  ][line width=0.75]    (10.93,-3.29) .. controls (6.95,-1.4) and (3.31,-0.3) .. (0,0) .. controls (3.31,0.3) and (6.95,1.4) .. (10.93,3.29)   ;
%Straight Lines [id:da3681075233854576] 
% Lines
\draw  (91.52,143) -- (91.52,381) ;
%Straight Lines [id:da13289399789061163] 
\draw    (91.52,195.55) -- (109.85,195.55) ;
\draw [shift={(111.85,195.55)}, rotate = 180] [color={rgb, 255:red, 0; green, 0; blue, 0 }  ][line width=0.75]    (10.93,-3.29) .. controls (6.95,-1.4) and (3.31,-0.3) .. (0,0) .. controls (3.31,0.3) and (6.95,1.4) .. (10.93,3.29)   ;
%Straight Lines [id:da4932753443830604] 
\draw    (91.52,296) -- (109.85,296) ;
\draw [shift={(111.85,296)}, rotate = 180] [color={rgb, 255:red, 0; green, 0; blue, 0 }  ][line width=0.75]    (10.93,-3.29) .. controls (6.95,-1.4) and (3.31,-0.3) .. (0,0) .. controls (3.31,0.3) and (6.95,1.4) .. (10.93,3.29)   ;
%Straight Lines [id:da4458194715081101] 
\draw    (91.52,381) -- (109.85,381) ;
\draw [shift={(111.85,381)}, rotate = 180] [color={rgb, 255:red, 0; green, 0; blue, 0 }  ][line width=0.75]    (10.93,-3.29) .. controls (6.95,-1.4) and (3.31,-0.3) .. (0,0) .. controls (3.31,0.3) and (6.95,1.4) .. (10.93,3.29)   ;
%Shape: Rectangle [id:dp846864727673416] 
% Element type
\draw   (248.83,102.82) -- (361,102.82) -- (361,143.77) -- (248.83,143.77) -- cycle ;
%Shape: Rectangle [id:dp5526144941829723]
% IP Type
\draw   (420.65,102.05) -- (533,102.05) -- (533,143) -- (420.65,143) -- cycle ;
%Shape: Rectangle [id:dp19401616889396212] 
%NVM-LUTs
\draw   (113.35,255) -- (237,255) -- (237,343) -- (113.35,343) -- cycle ;
%Shape: Rectangle [id:dp6447504381553044] 
% Others
\draw   (114.11,346.59) -- (237,346.59) -- (237,429) -- (114.11,429) -- cycle ;
%Shape: Rectangle [id:dp9560124537591201] 
% only LUTs
\draw   (274.43,165.64) -- (403.67,165.64) -- (403.67,252.27) -- (274.43,252.27) -- cycle ;
%Straight Lines [id:da4575698017894323] 
\draw    (254.1,143.77) -- (254.1,381.77) ;
%Straight Lines [id:da5181261072711036] 
\draw    (254.1,196.32) -- (272.43,196.32) ;
\draw [shift={(274.43,196.32)}, rotate = 180] [color={rgb, 255:red, 0; green, 0; blue, 0 }  ][line width=0.75]    (10.93,-3.29) .. controls (6.95,-1.4) and (3.31,-0.3) .. (0,0) .. controls (3.31,0.3) and (6.95,1.4) .. (10.93,3.29)   ;
%Straight Lines [id:da9554101252380751] 
\draw    (254.1,296.77) -- (272.43,296.77) ;
\draw [shift={(274.43,296.77)}, rotate = 180] [color={rgb, 255:red, 0; green, 0; blue, 0 }  ][line width=0.75]    (10.93,-3.29) .. controls (6.95,-1.4) and (3.31,-0.3) .. (0,0) .. controls (3.31,0.3) and (6.95,1.4) .. (10.93,3.29)   ;
%Straight Lines [id:da554873346457031] 
\draw    (254.1,381.77) -- (272.43,381.77) ;
\draw [shift={(274.43,381.77)}, rotate = 180] [color={rgb, 255:red, 0; green, 0; blue, 0 }  ][line width=0.75]    (10.93,-3.29) .. controls (6.95,-1.4) and (3.31,-0.3) .. (0,0) .. controls (3.31,0.3) and (6.95,1.4) .. (10.93,3.29)   ;
%Shape: Rectangle [id:dp8823561200329497] 
% only Switch boxes
\draw   (274.43,255.64) -- (404.67,255.64) -- (404.67,342.27) -- (274.43,342.27) -- cycle ;
%Shape: Rectangle [id:dp8161385948778308] 
%LUTs & switch boxes
\draw   (274.43,346.36) -- (404.67,345.36) -- (404.67,429) -- (274.43,429) -- cycle ;
%Shape: Rectangle [id:dp5278048823284927] 
% Soft IP
\draw   (446.25,165.86) -- (557,165.86) -- (557,252.5) -- (446.25,252.5) -- cycle ;
%Straight Lines [id:da7385288491182636] 
\draw    (425.92,143) -- (425.92,381) ;
%Straight Lines [id:da33801547250017494] 
\draw    (425.92,195.55) -- (444.25,195.55) ;
\draw [shift={(446.25,195.55)}, rotate = 180] [color={rgb, 255:red, 0; green, 0; blue, 0 }  ][line width=0.75]    (10.93,-3.29) .. controls (6.95,-1.4) and (3.31,-0.3) .. (0,0) .. controls (3.31,0.3) and (6.95,1.4) .. (10.93,3.29)   ;
%Straight Lines [id:da09099411816534575] 
\draw    (425.92,296) -- (444.25,296) ;
\draw [shift={(446.25,296)}, rotate = 180] [color={rgb, 255:red, 0; green, 0; blue, 0 }  ][line width=0.75]    (10.93,-3.29) .. controls (6.95,-1.4) and (3.31,-0.3) .. (0,0) .. controls (3.31,0.3) and (6.95,1.4) .. (10.93,3.29)   ;
%Straight Lines [id:da6732780847680595] 
\draw    (425.92,381) -- (444.25,381) ;
\draw [shift={(446.25,381)}, rotate = 180] [color={rgb, 255:red, 0; green, 0; blue, 0 }  ][line width=0.75]    (10.93,-3.29) .. controls (6.95,-1.4) and (3.31,-0.3) .. (0,0) .. controls (3.31,0.3) and (6.95,1.4) .. (10.93,3.29)   ;
%Shape: Rectangle [id:dp04244705373459734] 
% Firm IP
\draw   (446.25,255.86) -- (558,255.86) -- (558,341.9) -- (446.25,341.9) -- cycle ;
%Shape: Rectangle [id:dp8161156728940218] 
% Hard IP
\draw   (446.25,345.59) -- (558,345.59) -- (558,429) -- (446.25,429) -- cycle ;
%Image [id:dp2784685960143012] 
\draw (172.5,302.5) node  {\includegraphics[width=53.00pt,height=42.75pt]{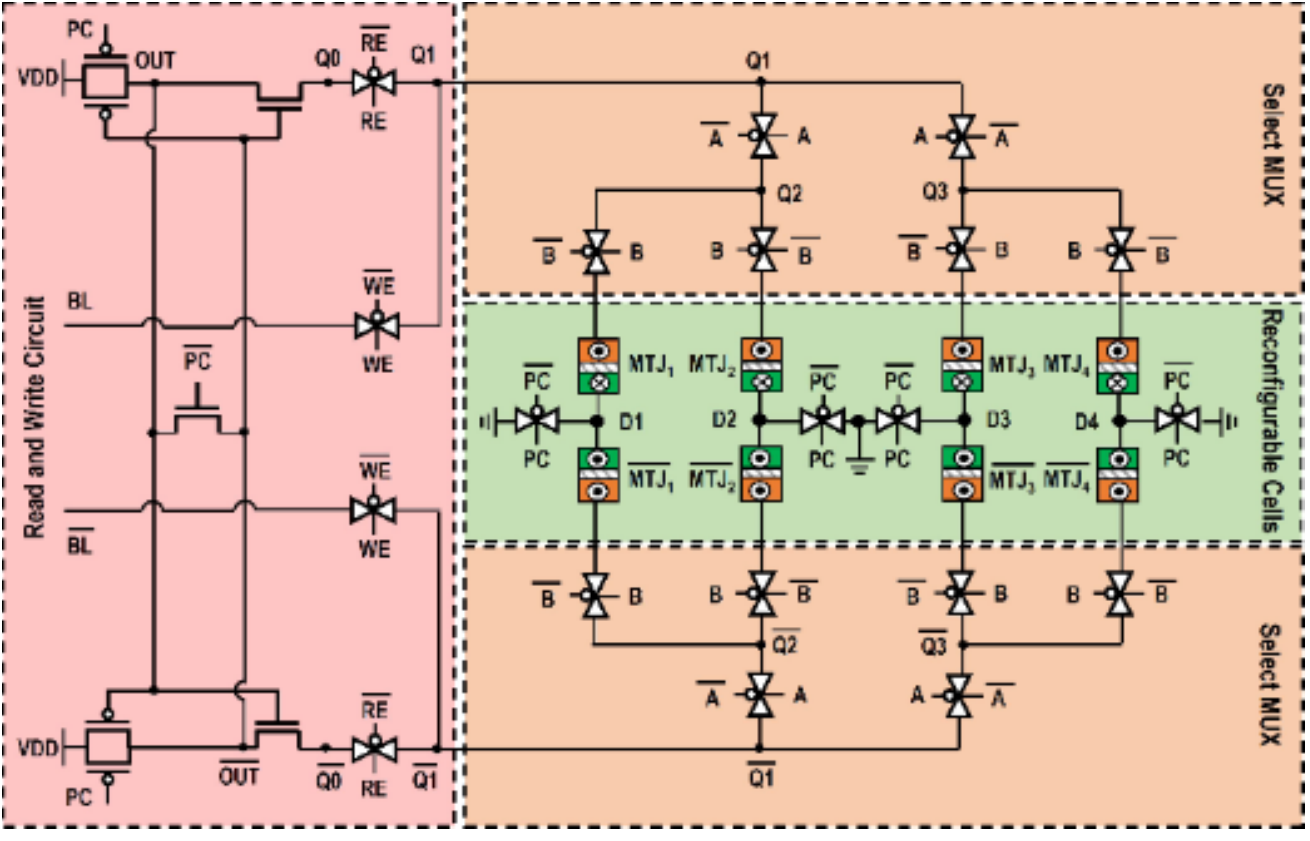}};
%Image [id:dp7084920306736378] 
\draw (172.5,392.5) node  {\includegraphics[width=53pt,height=41.75pt]{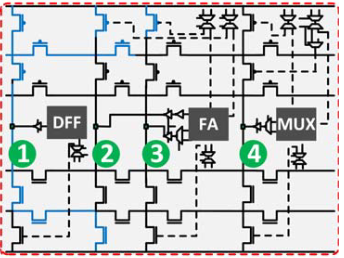}};
%Image [id:dp9518468659897166] 
\draw (338,215) node  {\includegraphics[width=50pt,height=50pt]{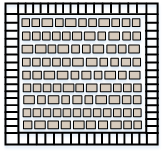}};
%Image [id:dp15876653786390404] 
\draw (338,303) node  {\includegraphics[width=48pt,height=48pt]{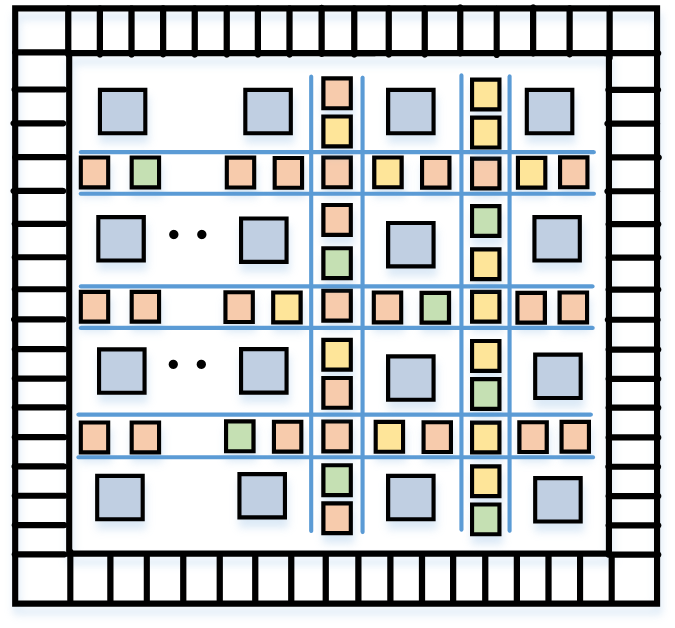}};
%Image [id:dp17077146930099585] 
\draw (338.215,395.17) node  {\includegraphics[width=47pt,height=47pt]{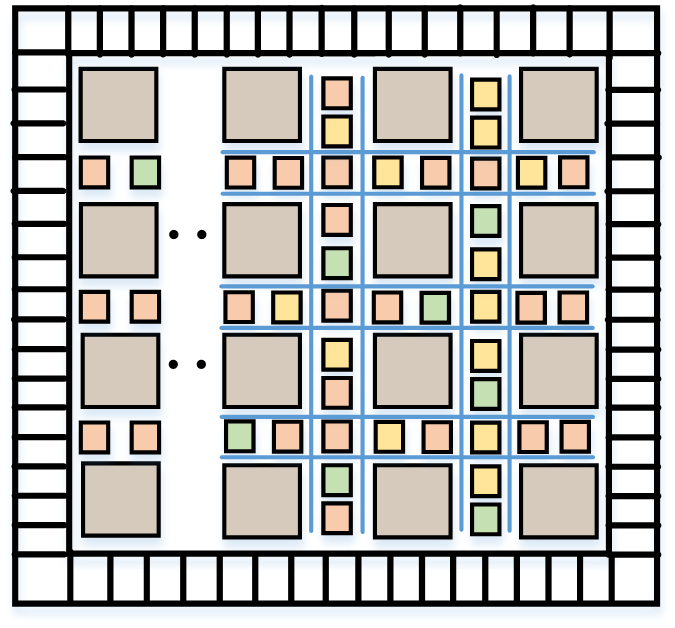}};
%Image [id:dp2284827149174804] 
\draw (500.17,214.17) node  {\includegraphics[width=75.00pt,height=35.00pt]{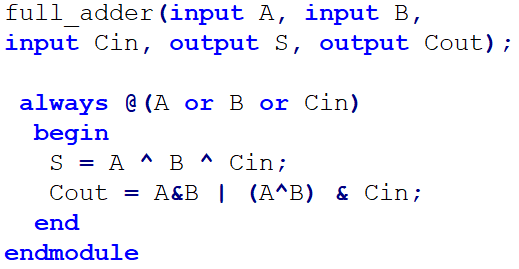}};
%Image [id:dp34093473291672005] 
\draw (502.17,301.17) node  {\includegraphics[width=70.75pt,height=40.75pt]{Figures/LL_a.pdf}};
%Image [id:dp06911622059976574] 
\draw (501,393.17) node  {\includegraphics[width=65.0pt,height=45.0pt]{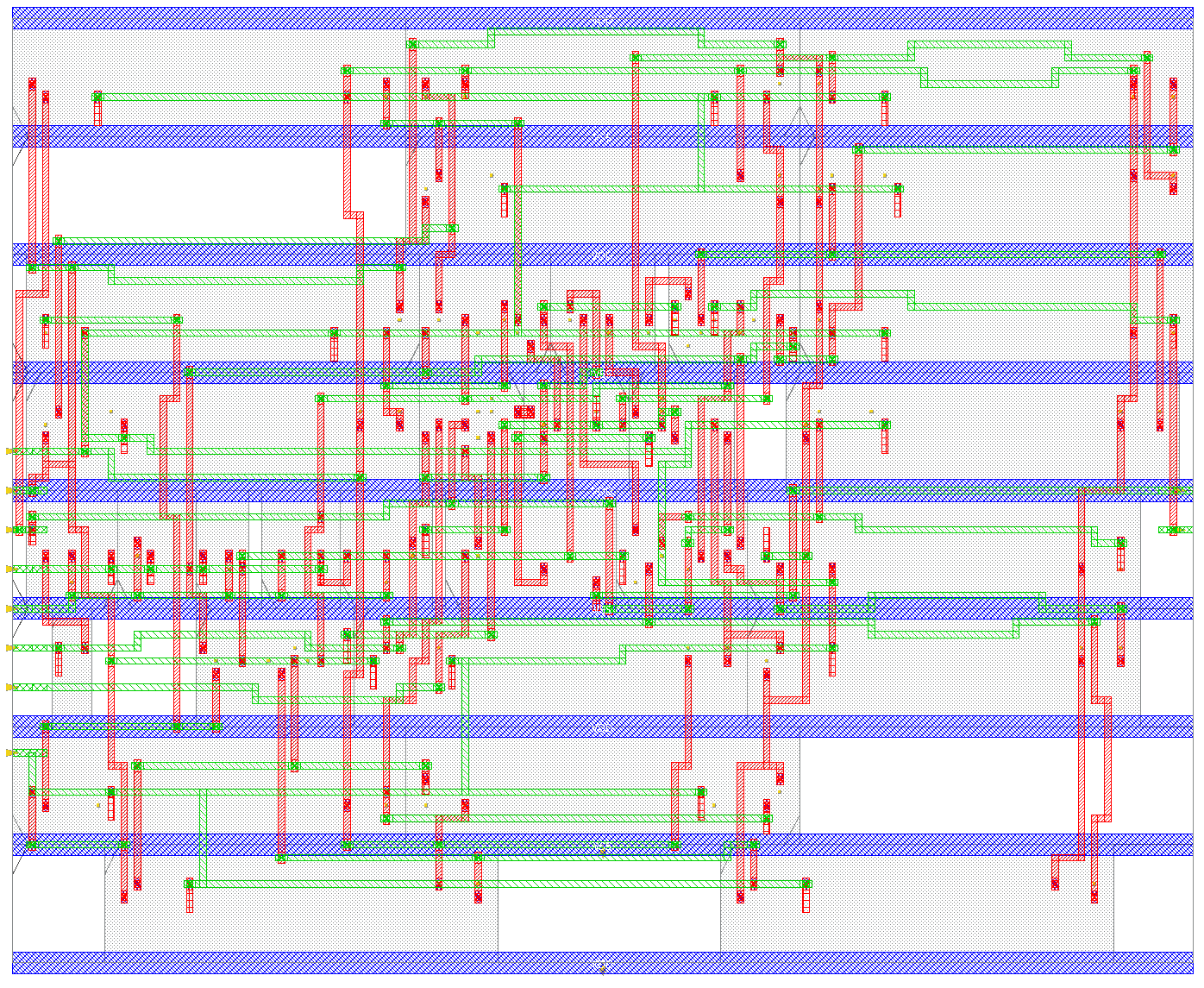}};
%Image [id:dp29301754441488415] 
\draw (168.67,215.33) node  {\includegraphics[width=36pt,height=50pt]{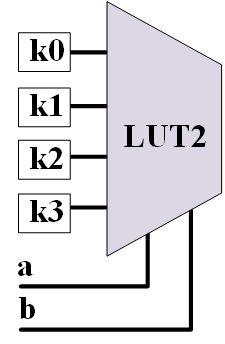}};

% Text Node
\draw (250.35,13.41) node [anchor=north west][inner sep=0.75pt]  [font=\small] [align=left] {\begin{minipage}[lt]{101.68pt}\setlength\topsep{0pt}
\begin{center}
\textbf{Reconfigurable-based }\\\textbf{Obfuscation}
\end{center}
 
\end{minipage}};
% Text Node
\draw (98.51,113.25) node [anchor=north west][inner sep=0.75pt]  [font=\small] [align=left] {\textbf{\textit{Technology Used}}};
% Text Node
\draw (266.97,112.48) node [anchor=north west][inner sep=0.75pt]  [font=\small] [align=left] {\textbf{\textit{Element Type}}};
% Text Node
\draw (457.41,112.25) node [anchor=north west][inner sep=0.75pt]  [font=\small] [align=left] {\textbf{\textit{IP Type}}};
% Text Node
\draw (139.28,166.57) node [anchor=north west][inner sep=0.75pt]  [font=\footnotesize] [align=left] {\textbf{\textit{{\footnotesize SRAM-LUTs }}}};
% Text Node
\draw (137.04,257.89) node [anchor=north west][inner sep=0.75pt]  [font=\footnotesize] [align=left] {\textbf{\textit{{\footnotesize NVM-LUTs \cite{kolhe2022lock} }}}};
% Text Node
\draw (147.54,347.8) node [anchor=north west][inner sep=0.75pt]  [font=\footnotesize] [align=left] {\textbf{\textit{{\footnotesize Others \cite{trap} }}}};
% Text Node
\draw (320.22,168.34) node [anchor=north west][inner sep=0.75pt]  [font=\footnotesize] [align=left] {\textbf{\textit{{\footnotesize LUTs}}}};
% Text Node
\draw (305.52,256.6) node [anchor=north west][inner sep=0.75pt]  [font=\footnotesize] [align=left] {\textbf{\textit{{\footnotesize Switch Boxes}}}};
% Text Node
\draw (290.16,347.59) node [anchor=north west][inner sep=0.75pt]  [font=\footnotesize] [align=left] {\textbf{\textit{{\footnotesize LUTs \& Switch Boxes}}}};
% Text Node
\draw (480.07,167.87) node [anchor=north west][inner sep=0.75pt]  [font=\footnotesize] [align=left] {\textbf{\textit{{\footnotesize Soft IP}}}};
% Text Node
\draw (480.33,257.34) node [anchor=north west][inner sep=0.75pt]  [font=\footnotesize] [align=left] {\textbf{\textit{{\footnotesize Firm IP }}}};
% Text Node
\draw (479.51,348.59) node [anchor=north west][inner sep=0.75pt]  [font=\footnotesize] [align=left] {\textbf{\textit{{\footnotesize Hard IP}}}};
\end{tikzpicture}
}
\caption{Classification of Reconfigurable-based Obfuscation}
\label{fig:classification}
\end{center}
\end{figure}

As discussed in the previous subsection, the research community has shown a growing interest in implementing  obfuscation using reconfigurable logic -- we can expect significant progress in this technique due to the apparent robust security guarantees it offers. It is thus important to adopt a unified classification and terminology the technique. As shown in Fig. \ref{fig:classification}, by carefully analyzing all the proposed techniques, we broadly classify them based on three important factors: 1) The technology used; 2) Element type; and 3) IP type.

%%%%%%%%%%%%%%%%%%%%%%%%%%%%%%%%%%%%%%%%%%%%%%%%%%%%%%%%%%%%%%%%%%%%%%%%%%%%%%%
%% FIGURE 
%%%%%%%%%%%%%%%%%%%%%%%%%%%%%%%%%%%%%%%%%%%%%%%%%%%%%%%%%%%%%%%%%%%%%%%%%%%%%%%

\subsubsection{Technology Used} \label{sec:eFPGA}
Numerous reconfigurable-based obfuscation techniques have been proposed to date. In a general sense, it is the LUT that actually enables logic to be obfuscated via reconfigurability. During the obfuscation process, the selected internal gates from the design are mapped onto  LUTs. As depicted in Fig. \ref{fig:classification}, there are several technologies available to store the key bits of the aforementioned LUTs. As shown in Fig. \ref{fig:classification}, we categorize them into SRAM-based LUTs \cite{reconfigure_1, reconfigure_3, kolhe2022silicon, lut1, chowdhury2021enhancing}, non-volatile memory (NVM)-based LUTs \cite{attaran2018static, winograd2016hybrid, yang2018exploiting, lut2, lut3, kolhe2021securing, lut5}, and Others \cite{eASIC, intro_epic, e-FPGA1, e-FPGA2, e-FPGA3, e-FPGA4, trap, Zain_TCAD,  mohan2021hardware, chen2021area, patnaik2018advancing, tomajoli2022alice, rangarajan2020opening, sathe2022investigating}. In Others, various technologies are available for programming the LUTs, including the spin transfer torque (STT)-based LUTs that exploit magnetic technology, flip-flop (FF)-based LUTs, programming pf individual transistors such as in a TRAnsistor-level Programming (TRAP) fabric, and programming of eFuses.

\begin {figure}[h!]
\centerline{\includegraphics[width=1.0\linewidth]{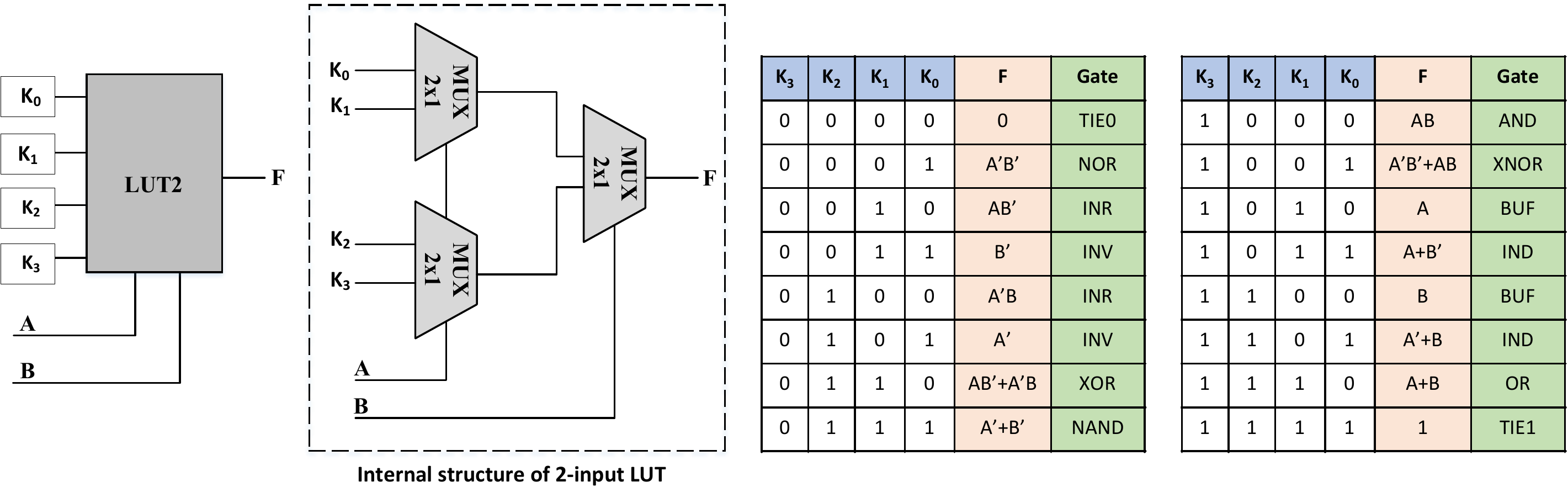}}
\caption{The 2-input LUT realizes all 2-input logic functions depending on the configuration bits K\textsubscript{0}-K\textsubscript{3}.}
\label{fig:gates_lut_example}
\end{figure}

Among these technologies, SRAM-based LUT has garnered significant attention to store the key bits. SRAM-based LUTs are often considered due to their programmability, reconfigurability, fast access time, low power consumption, smaller area, scalability, and ease of testing. These characteristics make them a preferred choice for implementing logic functions in FPGA designs and therefore a natural choice for obfuscation as well. Fig. \ref{fig:gates_lut_example} showcases an instance of a 2-input LUT and the various feasible functions it may implement. The 2-input LUT has the capacity to implement 16 distinct functions, as enumerated in the table presented in Fig. \ref{fig:gates_lut_example}.

In contrast to SRAM-based LUTs, the implementation of NVM-based LUTs relies on Non-Volatile Memory technology. The obvious advantage is that the programming remains even if the device is powered off. Compared to SRAM-based LUTs, however, they have some drawbacks, such as slower access time, and limited, complex programmability. However, it offers high-density storage elements. STT-based LUTs have been considered to design very robust and reverse engineering resilient LUTs. Contrary to NVM-based solutions, utilizing an FF-based LUT implementation renders the framework technology-agnostic, simplifying the process of floorplanning and placement tremendously. However, it does not achieve the same bit density as an SRAM-based solution does.

The TRAP fabric is a special case; it was introduced to obfuscate the design's intent by programming a sea of transistors. Efuses are essentially one-time programmable fuses that are blown to permanently program a specific configuration of the LUT. This differs from SRAM-based LUTs, which require reconfiguration each time the device powers up. Efuses offer different security properties, while they disallow reprogramming, they potentially expose the programmed values to reverse engineering by an end-user (but not by the foundry).

%%%%%%%%%%%%%%%%%%%%%%%%%%%%%%%%%%%%%%%%%%%%%%%%%%%%%%%%%%%%%%%%%%%%%%%%%%%%%%%
%% FIGURE 
%%%%%%%%%%%%%%%%%%%%%%%%%%%%%%%%%%%%%%%%%%%%%%%%%%%%%%%%%%%%%%%%%%%%%%%%%%%%%%%

\subsubsection{Element Type} \label{sec:LUT}
Concerning the element type, the reconfigurable-based obfuscation technique can leverage solely LUTs as in \cite{reconfigure_1, reconfigure_3, eASIC, kolhe2022silicon, lut1, Zain_TCAD, attaran2018static, winograd2016hybrid, yang2018exploiting, lut2, lut3, lut5}. The majority of the surveyed works take this approach, where the logic elements are obfuscated by LUT use only. As an example, the layout of this type of technique is illustrated in the top center of Fig. \ref{fig:classification}. This demonstrates a regular structure where the majority of the circuit consists of LUTs. On the other hand, there are some techniques that rely solely on exploiting switch boxes to obfuscate the design \cite{e-FPGA4, trap} by obfuscating connections between elements, as shown in the center of Fig. \ref{fig:classification}. Here, the layout is a combination of switch boxes and standard logic, which is indicated with blue color. Finally, the reconfigurable-based obfuscation techniques that utilize both LUTs and switch boxes are commonly referred to as FPGA redaction techniques \cite{intro_epic, e-FPGA1, e-FPGA2, e-FPGA3,  mohan2021hardware, chen2021area, patnaik2018advancing, tomajoli2022alice, rangarajan2020opening, sathe2022investigating, kolhe2021securing, chowdhury2021enhancing}. 

An example of reconfigurable-based obfuscation utilizing eFPGA is described in Section \ref{subsec:background}. It seems to be the case that obfuscating both LUTs and switch boxes has additional benefits in terms of enhancing the security of the design. By obscuring routing blocks, attackers cannot analyze routing patterns and infer functionality from them, thus offering another opportunity to thwart the otherwise powerful Boolean satisfiability (SAT)-based attacks. The reconfigurable-based obfuscation techniques can also be classified by the form they are delivered an intellectual property (IP), which we elaborate on in the next section.  %Hence, the mix of the two helps increase the number of Davis–Putnam–Logemann–Loveland (DPLL) calls \cite{5227064}, which the SAT solver has to prune for finding a key.

\subsubsection{IP Type} \label{sec:STTLUT}
Reconfigurable-based obfuscation techniques require that the designers perform several additional steps in their design flow. Selecting the critical modules to be redacted or selecting suitable gates to be replaced using LUTs are examples of additional steps. Hence, the process of obfuscation can be carried out at any of the three forms of IP, namely soft IP, firm IP, or hard IP, as illustrated in Fig. \ref{fig:classification}. Obfuscation carried out at the register transfer level (RTL), such as in Verilog or VHDL code, or any high-level codes can be classified as obfuscation at the soft IP level \cite{e-FPGA1, e-FPGA2, mohan2021hardware, chen2021area, patnaik2018advancing, rangarajan2020opening, sathe2022investigating}. The Verilog or VHDL IP is typically input by the designer and then modified in a way that results in design obfuscation. On the other hand, it can also be carried out with high-level codes, such as C/C++, followed by high-level synthesis. In this process, user-defined algorithms are executed to identify the optimal portions of modules to be obfuscated. 

The next category involves implementing obfuscation at the firm IP level, which is the most commonly utilized approach \cite{reconfigure_1, reconfigure_3, eASIC, kolhe2022silicon, lut1, Zain_TCAD, chowdhury2021enhancing, attaran2018static, winograd2016hybrid, yang2018exploiting, lut2, lut3, kolhe2021securing, lut5, intro_epic, e-FPGA3, trap, mohan2021hardware}. In this process, the post-synthesis netlist file is used as an input and it generates an obfuscated netlist as the output. 

Finally, the obfuscation can also be performed at the hard IP level where a crucial part of the design is identified and mapped into a hard IP. This entails using eFPGAs \cite{e-FPGA4, tomajoli2022alice} as the means to carry out the obfuscations, one of the examples is illustrated in Fig. \ref{fig:FPGA_redaction} and explained earlier at the end of Section \ref{subsec:background}.

%%%%%%%%%%%%%%%%%%%%%%%%%%%%%%%%%%%%%%%%%%%%%%%%%%%%%%%%%%%%%%%%%%%%%%%%%%%%%%%
%% SECTION - 3 (EXISTING RECONFIGURABLE-BASED OBFUSCATION)
%%%%%%%%%%%%%%%%%%%%%%%%%%%%%%%%%%%%%%%%%%%%%%%%%%%%%%%%%%%%%%%%%%%%%%%%%%%%%%%
\section{Existing Reconfigurable-based obfuscation techniques} \label{sec:Principles_com}
The section highlights a few of the most prominent reconfigurable-based defense techniques. It also presents a comparison among them. 

As the primary form of comparison among the various techniques, let us look into area, power, and delay overheads, as presented in Table \ref{tab:sec_3}. The majority of papers surveyed report their results considering PPA as a percentage increase over a baseline, which is also the approach we have followed. Table \ref{tab:sec_3} provides valuable insights into the impact of different obfuscation methods on key design metrics and can help guide the selection of an appropriate obfuscation technique based on specific design requirements/constraints. The techniques presented in Table \ref{tab:sec_3} are selected to highlight the extremes in PPA overheads. Moreover, the selected techniques also exhibit different flavors of obfuscation based on the technology used and element type. The results reported in Table \ref{tab:sec_3} represent the techniques that were compared with the baseline designs. Instead of highlighting the increase in PPA relative to baseline designs when reporting their results, the majority of these techniques simply focus on reporting the PPA of obfuscated designs.

Notice that Table \ref{tab:sec_3} is divided into two parts: CMOS and Emerging Technologies. As far as technology is concerned, the majority of the techniques that utilize CMOS-based LUTs are referred to as CMOS technology (CMOS TECH) in Table \ref{tab:sec_3}. However, emerging technologies that are used to implement LUTs, such as spin-transfer torque (STT), magnetic tunnel junction (MTJ), spin-orbit torque (SOT), and magnetic-random access memory (MRAM), are called emerging technologies (Emerging TECH). As a general rule of thumb, these two are very distinct from one another, therefore, trends and comparisons are more meaningful within the same technology class. 

It is noteworthy that the reconfigurable-based obfuscation  techniques using CMOS technology provide an analysis of large designs or benchmarks which highlights an increase in the PPA. However, most of the emerging technology-based techniques have only been evaluated on small design or ISCAS benchmarks, they exhibit a significant increase in PPA, as seen in references \cite{reconfigure_3, mohan2021hardware, eASIC, Zain_TCAD}. It is noticeable that some of the overheads are very large. For instance, in \cite{attaran2018static}, the authors reported a 95.06x increase in the area of the c2660 circuit from the ISCAS'85 benchmark suite. The work in \cite {reconfigure_1} suggests LUTs for obfuscation purposes and provides several replacement strategies to secure a netlist. %However, the paper does not address the topic of resilience against SAT-based attacks. 
On the other hand, \cite {reconfigure_3} utilizes an SRAM-LUT structure as configurable logic for gate replacement, incorporating a 2\textsuperscript{n}-to-1 MUX and 2\textsuperscript{n} configuration memory cells. This approach facilitates the dynamic configuration of the replaced gates. Nonetheless, using SRAM for logic obfuscation generally incurs a relatively high area overhead, as shown in Table \ref{tab:sec_3}. %Additionally, SRAM modules also have to be integrated during physical synthesis with careful consideration for their location and power delivery strategy. 
In \cite{e-FPGA4}, the authors have employed an eFPGA to redact the design and they considered the integration of different fabric sizes in PicoSoC. All three variants present a non-linear percentage increase in PPA with respect to the fabric size. A similar type of approach is employed in \cite{mohan2021hardware}, with substantially greater area and power overheads in comparison to other techniques. However, the delay overhead is significantly considerable. Additionally, it is essential to consider the security vs area and performance trade-offs, as given in \cite{Zain_TCAD}. 

%%%%%%%%%%%%%%%%%%%%%%%%%%%%%%%%%%%%%%%%%%%%%%%%%%%%%%%%%%%%%%%%%%%%%%%%%%%%%%%
%% TABLE 
%%%%%%%%%%%%%%%%%%%%%%%%%%%%%%%%%%%%%%%%%%%%%%%%%%%%%%%%%%%%%%%%%%%%%%%%%%%%%%%
\begingroup
\setlength{\tabcolsep}{2.0pt} % Column spacing - Default value: 6pt
\renewcommand{\arraystretch}{1.1} % Row spacing  - Default value: 1
\begin{table} [tb]
\footnotesize \centering
\caption{Comparison of Reconfigurable Obfuscation Techniques}
\label{tab:sec_3}
\begin{tabular}{|p{0.4cm}|p{3.5cm}|p{2.0cm}|p{1.5cm}|p{1.5cm}|p{1.5cm}|} \hline
                     % Header of the Table
                     & \textbf{Technique - Ref} & \textbf{Circuits} & \textbf{Area (\%)} & \textbf{Power (\%)} & \textbf{Delay (\%)} \\ \cline{2-6} 
                     % Data inserted here
                     % 1
                     & \multirow{2}{*}{eRECONF LOGIC - \cite{reconfigure_3}} & IDU & 1595.0 & 942.8 & 165.0 \\
                     &  & LEON2 & 34.7 & 6.7 & 131.0 \\ \cline{2-6} 
                     % 2
                     & \multirow{4}{*}{eFPGA REDAC - \cite{e-FPGA4}}  & PicoSoC + 3×3 & 10.0 & 30.0 & 50.0 \\
                     & & PicoSoC + 4×4 & 30.0 & 60.0 & 80.0  \\
                     & & PicoSoC + 5×5 & 60.0 & 90.0 & 200.0 \\
                     & & PicoSoC + 6×6 & 140.0 & 130.0 & 270.0 \\ \cline{2-6} 
                     % 3
                     & \multirow{2}{*}{FINE-GRAINED eFPGA - \cite{mohan2021hardware} } & RISC-V & 89.0 & 40.0 & 136.0 \\
                     & & GPS & 39.0 & 46.0 & 0.0 \\ \cline{2-6} 
                     % 4
                     & hASIC - \cite{eASIC} & SHA-256 & 2231.8 & 717.0 & 353.8 \\ \cline{2-6} 
                     & CAD-hASIC - \cite{Zain_TCAD} & SHA-256 & 4192.0 & 2051.9 & 150.0 \\ \cline{2-6}
                     % 5 
\multirow{-12}{*}{\rotatebox[origin=c]{90}{\textbf{CMOS TECH}}} 
                    % tech
                    & SILICON-LUT - \cite{kolhe2022silicon} $\dagger$ & Multiple & \begin{tabular}[c]{@{}l@{}}LO: 7.0\\ MO: 14.0\\ HO: 262.0\end{tabular} & \begin{tabular}[c]{@{}l@{}}LO: 0.0\\ MO: 3.5\\ HO: 17.8\end{tabular} & 0.0 \\ \hline
                   % tech
                   & Hybrid STT-LUT - \cite{winograd2016hybrid} & ISCAS & \begin{tabular}[c]{@{}l@{}}min: 0.1 \\ avg: 6.4 \\ max: 20.6\end{tabular} & \begin{tabular}[c]{@{}l@{}}min: 0.7 \\ avg: 24.96 \\ max: 82.11\end{tabular} & \begin{tabular}[c]{@{}l@{}}min: 0.0 \\ avg: 28.4 \\ max: 82.3\end{tabular} \\ \cline{2-6}  
                  % tech
                  & \multirow{7}{*}{MTJ-STT-LUT - \cite{lut5} }  & c2670 & 91.5 & 53.3 & 0.0 \\ 
                  & & c7552 & 91.5 & 20.4 & 0.0 \\ & & B12  & 60.5 & 18.5 & 0.0 \\
                  & & FIR & 43.1 & 17.3 & 0.0 \\
                  & & IIR & 8.4 & 10.1 & 0.0 \\
                  & & AES & 4.9 & 2.8 & 0.0 \\
                  & & DES & 3.3 & 2.5 & 0.0 \\ \cline{2-6}
                  % tech
                  & SOT-LUT-16i\_{G} - \cite{yang2018exploiting} & ISCAS/MCNC & \begin{tabular}[c]{@{}l@{}}min: 2.5 \\ avg: 12.2 \\ max: 25.4\end{tabular}  & -- & \begin{tabular}[c]{@{}l@{}}min: 0.0 \\ avg: 20.4 \\ max: 41.8\end{tabular} \\ \cline{2-6} 
                   % tech
                  & SOT-LUT-32i\_{G} - \cite{yang2018exploiting} & ISCAS/MCNC & \begin{tabular}[c]{@{}l@{}}min: 6.7 \\ avg: 17.7 \\ max: 27.2\end{tabular} & -- & \begin{tabular}[c]{@{}l@{}}min: 0.0 \\ avg: 28.5\\ max: 47.5\end{tabular} \\ \cline{2-6} 
                  % tech
                  & SOT-LUT-64i\_{G} - \cite{yang2018exploiting} & ISCAS/MCNC & \begin{tabular}[c]{@{}l@{}}min: 15.2 \\ avg: 22.2 \\ max: 27.2\end{tabular} & -- & \begin{tabular}[c]{@{}l@{}}min: 4.2 \\ avg: 36.1\\ max: 78.2\end{tabular} \\ \cline{2-6}
                  % tech
                  & \multirow{6}{*}{CGRRA - \cite{chen2021area} } & sort & 193.0 & -- & 70.0 \\
                  & & cordic & 492.0 & -- & 66.0 \\
                  & & interp & 432.0 & -- & 170.0 \\
                  & & decim & 147.0 & -- & 63.0 \\ 
                  & & fft & 861.0 & -- & 34.0 \\ 
                  & & cnn & 7.0 & -- & 45.0 \\ \cline{2-6} 
                  % tech
                  & \multirow{3}{*}{TRAP - \cite{e-FPGA3} } & AMT & 4.0 & 0.2 & 6.0 \\
                  & & AMT+RSR+BP & 9.0 & 0.2 & 83.0 \\
\multirow{-19}{*}{\rotatebox[origin=c]{90}{\textbf{Emerging TECH}}} & & Dispatch  & 20.0 & 0.6 & 164.0 \\ \hline
\end{tabular}
\begin{tablenotes}
      \centering
      \item  \footnotesize $\dagger$ Low-obfuscation (LO), Medium-obfuscation (MO), High-obfuscation (HO).  
    \end{tablenotes}
%\vspace{-2mm}
\end{table}
\endgroup

%Additionally, the power side-channel signature of SRAM-based LUTs enables the threat of power side-channel attacks (P-SCAs). Hence, the challenges in the field of SRAM-LUT based reconfigurable obfuscation necessitate an innovative approach that can resist SAT-attack while being resilient to P-SCA.

\begin {figure}[h!]
\centerline{\includegraphics[width=0.75\linewidth]{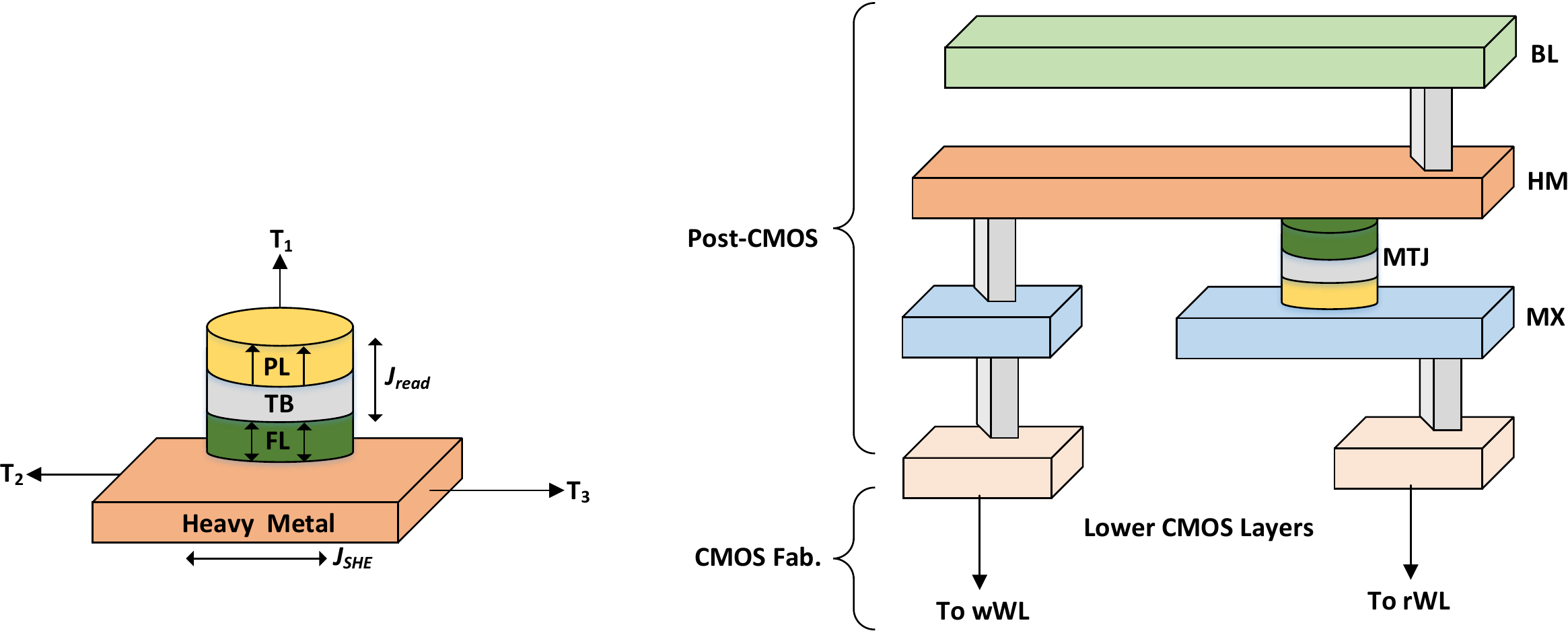}}
\caption{Physical structure demonstration of SHE assisted MTJ switching mechanism (adapted from \cite{yang2018exploiting}).}
\label{fig:STT_MJT_SOT}
\end{figure}

Almost all the techniques aim to redact the most sensitive part of the design in a modular approach. The flexibility to obfuscate any part of the design, thereby crossing module boundaries, is presented in \cite{eASIC, Zain_TCAD}. The authors almost obfuscated the majority of the gates in the design, approximately 80-90\%, which then yields high PPA penalties. %Recalling again the emerging technologies for reconfigurable-based obfuscation techniques include STT, MTJ, SOT, MRAM, and transistor-level configuration.  A similar strategy could be employed for the combination of both switch box and LUTs, where the LUT is designed using a hybrid of CMOS and emerging technology.
Every technique that lies in the category of emerging technology could be referred to as a hybrid because they utilize standard CMOS technology along with one of the STT, MTJ, SOT, or MRAM devices. Table \ref{tab:sec_3} provides an overview of the techniques that belong to these technologies, LUT being used as an element type designed using a hybrid of CMOS and emerging technology. A similar strategy could be employed for the combination of both switch box and LUTs as a hybrid technology. To further elaborate, we would like to emphasize a few insights about the hybrid technologies mentioned earlier. In order to realize a Spin Transfer Torque (STT) device, stacked multilayer sandwich structures are typically used. According to Fig. \ref{fig:STT_MJT_SOT}, there are a number of layers in the structure, including an oxide tunnel barrier, a free magnetic layer, and a pinned magnetic layer. An external magnetic field or a spin-polarized current J\textsubscript{read} flowing through the junction can switch the magnetization direction of the free layer from a parallel to an antiparallel state (P to AP). These states, in turn, can represent a logic-1 or a logic-0. However, the most important aspect of the device depicted in \ref{fig:STT_MJT_SOT} is that the MTJ itself does not compete with standard cells for area since it resides between two metal layers.

%The switching mechanism occurs in-plane when the current density exceeds a critical value, J\textsubscript{c}, which is as low as $8 \times 105 A/cm^2$ in CoFeB/MgO/CoFeB stack structures \cite{MJT_driven}. Since the spin-MTJ device surface area is typically small (e.g., $113 nm \times 75 nm$ or less), the critical current is less than 100 $\mu A$ and can be generated by a typical CMOS current source. Additionally, J\textsubscript{c} can be reduced by integrating perpendicular magnetic anisotropy (PMA) in the free layer of the MTJ, so that stable magnetization points out-of-plane instead of within the same plane as the free layer \cite{Ikeda2010}. 

%%%%%%%%%%%%%%%%%%%%%%%%%%%%%%%%%%%%%%%%%%%%%%%%%%%%%%%%%%%%%%%%%%%%%%%%%%%%%%%
%% FIGURE 
%%%%%%%%%%%%%%%%%%%%%%%%%%%%%%%%%%%%%%%%%%%%%%%%%%%%%%%%%%%%%%%%%%%%%%%%%%%%%%%

\begin {figure}[h!]
\centerline{\includegraphics[width=0.55\linewidth]{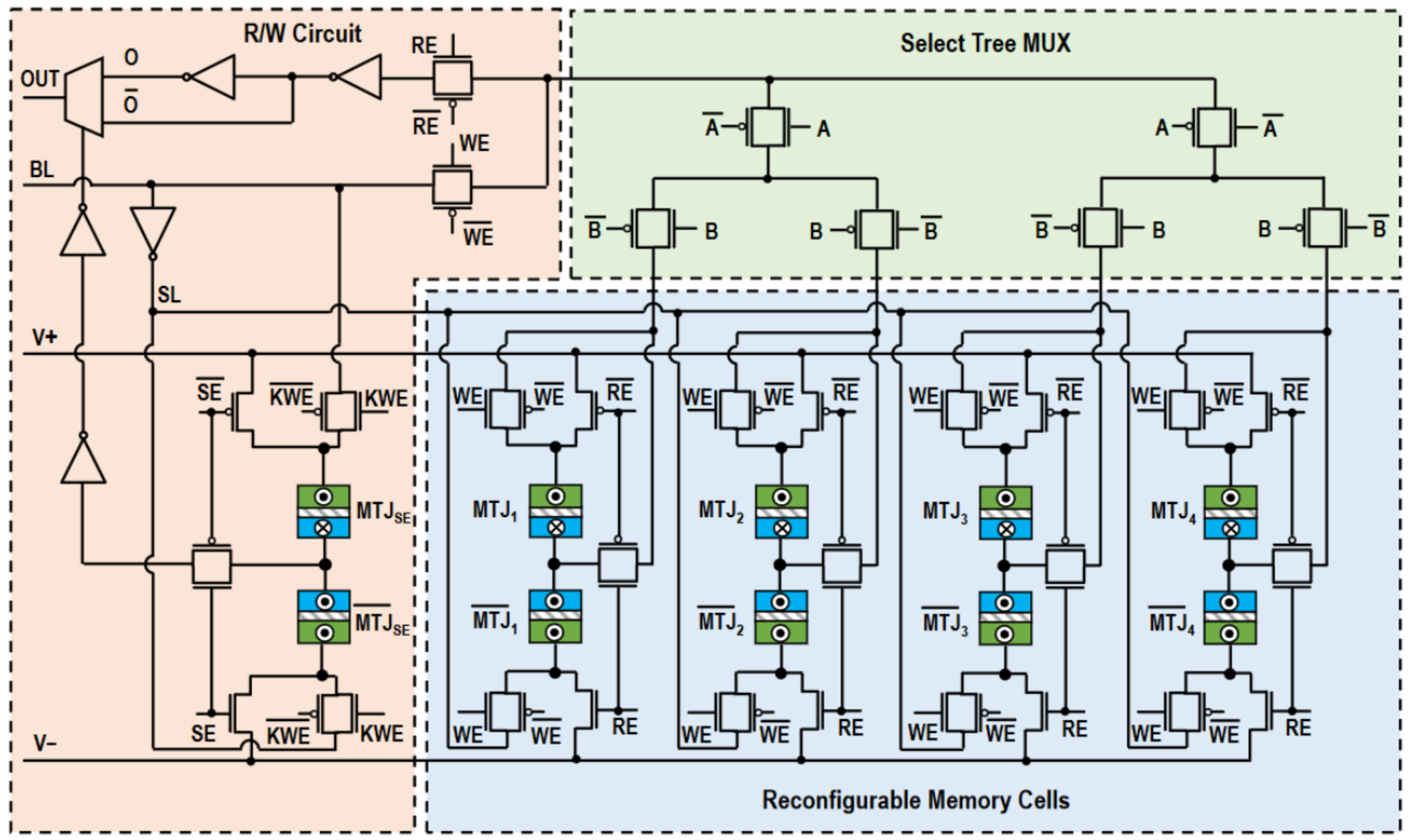}}
\caption{The circuit diagram of the proposed 2-input MRAM-based LUT) that utilizes STT-MTJ devices \cite{kolhe2021securing}}
\label{fig:MRAM_LUT}
\end{figure}

MTJ integration requires little die area, except for the CMOS circuits and contacts used to connect MTJs to MOS transistors. The MTJs does come with certain operational challenges. Asymmetry in write and read operations results in a difference in operation energy and delay, requiring a higher current for completing write operations. Spin-orbit interaction has recently been explored as an alternative write approach to overcome these bottlenecks, details area available from \cite{SOT_explain}. The authors in \cite{kolhe2021securing} proposed reconfigurable logic and interconnects (RIL)-Blocks that leverage MRAM-based LUTs with MTJ technology and routing-based obfuscation. As illustrated in Fig. \ref{fig:MRAM_LUT}, each cell is accessed via A and B, while the write operation is controlled by $\overline{WE}$ signals. In each memory cell, the MTJs change complementary to each other during write operations. According to input signals A and B, output nodes O and O direct the appropriate output to the select tree MUX using the RE and RE signals. The RIL-Blocks are built using commercially available STT-MTJ technology to provide the desired obfuscation.

%%%%%%%%%%%%%%%%%%%%%%%%%%%%%%%%%%%%%%%%%%%%%%%%%%%%%%%%%%%%%%%%%%%%%%%%%%%%%%%
%% FIGURE 
%%%%%%%%%%%%%%%%%%%%%%%%%%%%%%%%%%%%%%%%%%%%%%%%%%%%%%%%%%%%%%%%%%%%%%%%%%%%%%%

%Let us assume an example of SOT-based LUT, the obfuscated circuits fabricated in this way require fewer masks to integrate the spin hall effect (SHE)-MTJ during the backend process \cite{yang2018exploiting}, resulting in a lower total cost than the obfuscated circuits which exploit SRAM-based LUTs \cite{yang2018exploiting}. 
The technique in \cite{winograd2016hybrid} is the first one to propose hybrid-STT LUTs and its area overhead is pretty small but the power and delay overheads are around 82\%. The approach given in \cite{yang2018exploiting} involves examining the internal gates of each class to identify the gate that can be obfuscated with minimal design overhead and path delay. In this approach, the authors present a cluster of 16, 32, or logic gates which are replaced with 2, 3, and 4-input LUTs. While the area of this technique is comparable to \cite{winograd2016hybrid}, its maximum overhead is nearly equal for groups of 16, 32, and 64. The approach in \cite{lut5} does not incur any performance overhead and it exhibits almost 90\% increase in area. The method given in \cite{chen2021area} also validates their technique on small circuits but the area overhead is large even for a very small circuit that executes a sorting algorithm. The PPA overheads in \cite{trap, e-FPGA3} are very small, this approach has the lowest PPA in reconfigurable-based obfuscation with the same remark of validation on small circuits.

% CMOS technologies
Now, we will briefly explore how CMOS technologies can be used to obfuscate circuits. LUT-based obfuscation is the dominant technique used in CMOS technologies. In \cite {kolhe2021securing}, the authors presented a novel approach to thwart various types of attacks by utilizing both reconfigurable interconnect and logic blocks. The researchers presented their security analysis without highlighting PPA overheads. The authors in \cite{tomajoli2022alice} proposed RTL-based partitioning in tandem with eFPGA redaction for better obfuscation. However, the authors in \cite{e-FPGA1, e-FPGA2, sathe2022investigating} proposed a very similar partitioning scheme at the behavioral level, and demonstrated an automated partitioning flow for behavioral descriptions for high-level synthesis (HLS). This technique is associated with high PPA overheads and the explanation of this technique is detailed in Section \ref{sec:background}. A similar approach was implemented in \cite{mohan2021hardware} to obfuscate portions of the RISC-V control path and its overhead is mentioned in Table \ref{tab:sec_3}. In \cite {chowdhury2021enhancing}, the authors presented a LUT-based obfuscation algorithm that optimizes the replacement locations of LUTs and the input signals they receive in order to achieve resiliency against SAT attacks while minimizing overhead. The authors of \cite{lut1} have proposed a different scheme called LUT-Lock that focuses the effort on a minimal set of primary output pins to increase the obfuscation difficulty. To accomplish this, they decided to focus on fan-in with a few primary outputs for obfuscation and selected gates that are connected to the smallest number of output pins. Additionally, gates that exhibit less control over primary inputs are preferable for obfuscation. %The important feature is that avoiding back-to-back LUT replacement schemes considerably reduces the number of valid key possibilities, thereby increasing the resiliency of the proposed algorithm against SAT attacks. It enhances the number of calls in SAT-solvers, especially, Davis–Putnam–Logemann–Loveland (DPLL) calls, that the SAT-solver has to prune for finding a key. 

As the research on this type of techniques has matured, authors started to consider the trade-off space between security and PPA. The research in \cite {lut3} has demonstrated that circuits with a small LUT input size (e.g., 2-input LUT) can be easily de-obfuscated. Their research demonstrates that the input size of LUT is the most influential and straightforward factor in achieving SAT resiliency. Though LUT-based obfuscation provides high-security levels, it results in prohibitive PPA overhead \cite{reconfigure_3}. Recently, another LUT-based research work in \cite{Zain_TCAD} presents a security-aware computer-aided design (CAD) flow, which utilizes a combination of MUXs, LUTs, and FFs for obfuscation. It is compatible with the standard cell-based physical synthesis flow. The approach explores a midpoint between pure FPGA and pure ASIC design to generate heavily obfuscated designs that combine static parts with reconfigurable parts, which they them term a “hybrid ASIC” (hASIC). The results illustrate that better obfuscation could be achieved with slightly high hardware overhead.

Interestingly, the authors in \cite{kolhe2022silicon} demonstrated a low overhead digital IC design obfuscation flow which is compatible with existing Electronic Design Automation (EDA) tools. They have also fabricated an IC to prove the concept. The proposed method was demonstrated for both non-volatile internal (efuse) and volatile external (SRAM) LUT key configurations to showcase its flexibility. Using the fabricated silicon, the actual design overhead, in terms of area, performance, and power, was measured. The chip was evaluated for various levels of security and showed that the SAT attack could be thwarted in the low obfuscation case with minimal overhead and in the medium obfuscation case with a maximum of 14\% overhead, allowing for a significant SAT runtime margin.

%%%%%%%%%%%%%%%%%%%%%%%%%%%%%%%%%%%%%%%%%%%%%%%%%%%%%%%%%%%%%%%%%%%%%%%%%%%%%%%
%% FIGURE 
%%%%%%%%%%%%%%%%%%%%%%%%%%%%%%%%%%%%%%%%%%%%%%%%%%%%%%%%%%%%%%%%%%%%%%%%%%%%%%%
\begin {figure}[h!]
\centerline{\includegraphics[width=0.45\linewidth]{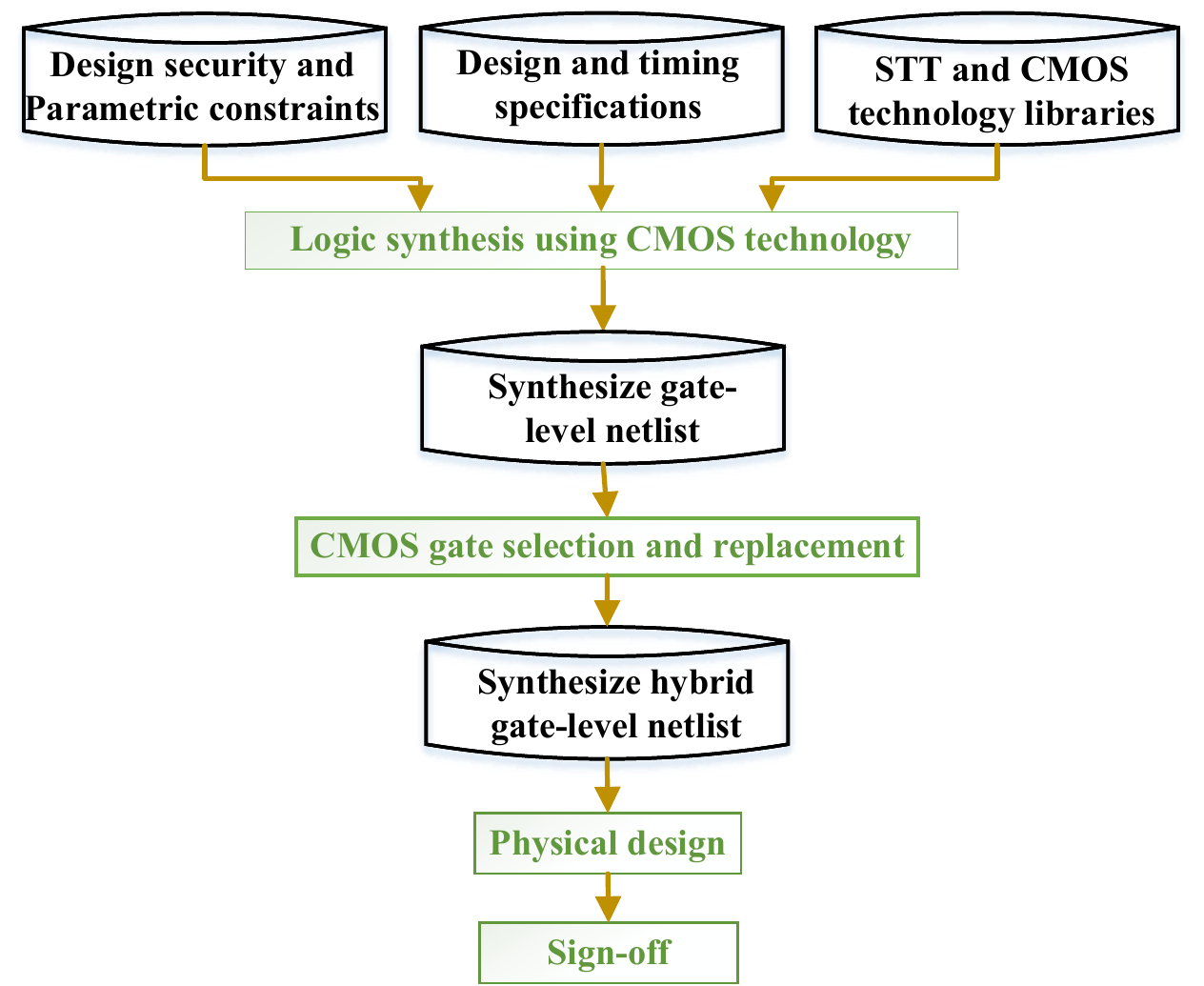}}
\caption{Security-driven hybrid STT-CMOS design flow (adapted from \cite{winograd2016hybrid})}
\label{fig:STTCMOS}
\end{figure}

\begin {figure}[h!]
\centerline{\includegraphics[width=0.45\linewidth]{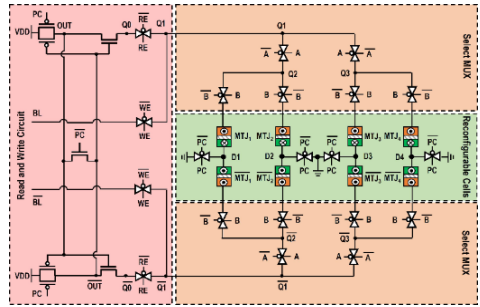}}
\caption{The circuit-level diagram of 2-input SyM-LUT using STT-MTJ devices \cite{kolhe2022lock}}
\label{fig:MRAMMJT}
\end{figure}

% Emerging technologies
Now, we will briefly explore how emerging technologies can be used to obfuscate circuits. In \cite {lut2}, the authors studied the previously proposed LUT-based obfuscation schemes and proposed a customized STT-LUT-based obfuscation with two different variants: LUT+MUX-based obfuscation, and LUT+LUT-based obfuscation. This combination assists them in elevating security provided by logic obfuscation while concurrently creating SAT-hard instances. But, the work in \cite {lut5} explored the design space for four crucial design factors that impact the design overhead and security of hybrid STT-LUT-based obfuscation: (1) LUT technology, (2) LUT size, (3) number of LUTs, and (4) replacement strategy as illustrated in Fig. \ref{fig:STTCMOS}. It is concluded in \cite {lut5} that, among the four studied parameters, the input size of LUT is the most influential and straightforward factor in achieving SAT resiliency. The authors in \cite {kolhe2022lock} proposed a multi-layer defense mechanism using a combination of a Symmetrical MRAM-based LUT (SyM-LUT) and STT-MTJ-based LUT as shown in Fig. \ref{fig:MRAMMJT} \cite{winograd2016hybrid}. SST structures require a high write current to switch the magnetization direction, SOT structures can achieve the same result with significantly reduced current. This makes them a promising candidate for low-power and high-speed data storage and processing applications. Therefore, the authors in \cite {yang2018exploiting} utilized the hybrid SOT-CMOS circuits to realize the reconfigurable logic with a lower write current among LUTs programming operations with smaller hardware overhead as shown in Table \ref{tab:sec_3}. As an example, Fig. \ref{fig:STTCMOS} illustrates the STT-CMOS hybrid design flow. In order to incorporate hybrid technologies into the design flow, additional parameters and processes must be considered. This approach differs from conventional design flows in that it includes additional steps, such as replacing gates and synthesizing the netlist again.  %This approach offers resiliency against various attacks such as state-of-the-art SAT-based attacks, removal attacks, and ML-assisted P-SCA with only a small overhead.

%%%%%%%%%%%%%%%%%%%%%%%%%%%%%%%%%%%%%%%%%%%%%%%%%%%%%%%%%%%%%%%%%%%%%%%%%%%%%%%
%% FIGURE 
%%%%%%%%%%%%%%%%%%%%%%%%%%%%%%%%%%%%%%%%%%%%%%%%%%%%%%%%%%%%%%%%%%%%%%%%%%%%%%%

The research community has modified the research aspect of reconfigurable domains by incorporating dynamic morphing to resist SAT attacks. The Giant Spin Hall Effect (GSHE) \cite{patnaik2018advancing} and Magneto-Electric Spin-Orbit (MESO) \cite{rangarajan2020opening} concepts enable polymorphism which allows for runtime configuration of Boolean functions. Dynamically changing between states during runtime offers obfuscation, hence neutralizing the threat of a SAT attack. However, randomly morphing from one state to another limits the applicability of the obfuscation to  applications like image processing that can tolerate a certain degree of errors \cite{rangarajan2020opening}. %Moreover, this defeats the purpose of obfuscation, as the attacker can deliberately fix the functionality of the MESO devices, and the IP may still function correctly since the application can tolerate a certain level of error. 
In \cite{chen2021area}, a variant of an existing architecture called coarse-grained runtime reconfigurable array (CGRRA) is used to selectively map different portions of the design into a reconfigurable block. Different portions of a design, executed at different clock cycles, can therefore be mapped onto the same CGRRA without requiring additional area. The overall hardware overhead of this scheme is mentioned in Table \ref{tab:sec_3}. Examples of this architecture include the Stream Transpose Processor from Renesas Electronics \cite{stprenesas} and Samsung’s Reconfigurable Processor \cite{kim2014ulp}. Another approach that offers reconfiguration at the transistor level is presented in \cite{trap}. The solution exploits ``sea-of-transistor'' architecture, supporting the implementation of custom cell libraries and facilitating fabric time-sharing. Here, the authors present a partitioning flow for RTL descriptions. As shown in Table \ref{tab:sec_3}, the results are promising to note that this solution not only results in an order of magnitude smaller PPA overhead but also increases the complexity of the design \cite{trap}. However, this approach presents some drawbacks: it can make testing and simulation more challenging than other solutions, and since the configuration is done at the transistor level, the resulting bitstream size will become extremely large (see Fig. 1 of \cite{e-FPGA3}). %A well-known factor is that the state-of-the-art test methods are developed either for ASICs or for FPGAs, unfortunately, it does not support this new transistor-level fabrics; and hence, to address this shortcoming, a novel application-agnostic test scheme specifically framed to the TRAP architecture proposed in \cite{trap}. 

%%%%%%%%%%%%%%%%%%%%%%%%%%%%%%%%%%%%%%%%%%%%%%%%%%%%%%%%%%%%%%%%%%%%%%%%%%%%%%%
%% FIGURE 
%%%%%%%%%%%%%%%%%%%%%%%%%%%%%%%%%%%%%%%%%%%%%%%%%%%%%%%%%%%%%%%%%%%%%%%%%%%%%%%
% \begin {figure}[h!]
% \centerline{\includegraphics[width=1.0\linewidth]{Figures/TRAPresult.pdf}}
% \caption{Overhead incurred by TRAP-based obfuscation of FabScalar modules and comparison to conventional FPGAs \cite{e-FPGA3} }
% \label{fig: TRAPresult}
% \end{figure}

Overall, the majority of reconfigurable-based techniques leverage either a reconfigurable element or the eFPGA redaction as depicted in the right panel of Fig. \ref{fig:FPGA_redaction}. The left panel of Fig. \ref{fig:FPGA_redaction} represents the ASIC that requires obfuscation, while the center panel demonstrates the LUT-based obfuscation approach utilized for obfuscation. The eFPGA-based obfuscation is explained in Section \ref{subsec:background}. In this approach, one of the crucial modules is transformed into reconfigurable logic, and the other modules are converted into standard cells. Additionally, the reconfigurable elements are dispersed throughout the layout, whereas the eFPGA macro is placed in a specific place, resulting in a concentrated reconfigurable logic. 

% %%%%%%%%%%%%%%%%%%%%%%%%%%%%%%%%%%%%%%%%%%%%%%%%%%%%%%%%%%%%%%%%%%%%%%%%%%%%%%%
% %% FIGURE 
% %%%%%%%%%%%%%%%%%%%%%%%%%%%%%%%%%%%%%%%%%%%%%%%%%%%%%%%%%%%%%%%%%%%%%%%%%%%%%%%
% \begin {figure}[h!]
% \centerline{\includegraphics[width=0.75\linewidth]{Figures/Comparison_Obfuscation_Redaction_vs_LUTs.pdf}}
% \caption{The comparison of two major reconfigurable-based obfuscation techniques. The individual panels are not to scale.}
% \label{fig:RE_obs_2}
% \end{figure}

%%%%%%%%%%%%%%%%%%%%%%%%%%%%%%%%%%%%%%%%%%%%%%%%%%%%%%%%%%%%%%%%%%%%%%%%%%%%%%%
%% Table
%%%%%%%%%%%%%%%%%%%%%%%%%%%%%%%%%%%%%%%%%%%%%%%%%%%%%%%%%%%%%%%%%%%%%%%%%%%%%%%
\section{Security Analysis: Threat Models and Existing Attacks} \label{sec:securit_analysis}
This section presents an overview of state-of-the-art attacks and their associated threat models. Additionally, a comparison of different attacks and a thorough security analysis are presented. Furthermore, this section provides a comparative analysis of logic locking and reconfigurable-based obfuscation techniques. 

\subsection{Threat models} \label{sec:threat_models}

\begin {figure}[h!]
\centerline{\includegraphics[width=0.40\linewidth]{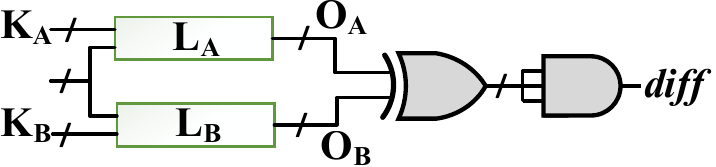}}
\caption{The Miter circuit, which is utilized in the SAT attack to identify DIPs (adapted from \cite{eval_logic}).}
\label{fig:miter}
\end{figure}

In the context of adversarial modeling, threat models can be classified into oracle-guided and oracle-less. This terminology has been borrowed from the field of Logic Locking, and it is equally applicable to attacks on reconfigurable-based obfuscation techniques. An oracle-guided attack involves a reverse-engineered netlist and a functional chip, which is commonly referred to as an oracle. As depicted in Figure \ref{fig:miter}, oracle-guided attacks typically employ Boolean satisfiability (SAT) techniques based on two copies of the locked netlist as part of a miter-like circuit. %Assume a functional IC receives an input $I\textsubscript{d}$ and produces a correct output $O\textsubscript{d}$. It is expected that both circuits (L\textsubscript{A} and L\textsubscript{B}) will produce different outputs under these circumstances. It is only possible to generate an output that is consistent with $O\textsubscript{d}$ for one of the two key values, whereas the other key value produces erroneous output for at least one input pattern. As $I\textsubscript{d}$ distinguishes the key search space into classes of feasible and infeasible inputs, it is referred to as a distinguishing input pattern (DIP). DIPs constitute the foundation of the SAT attack, which is essentially an iterative construction of DIPs as constraints until no further DIPs can be found. 

%%%%%%%%%%%%%%%%%%%%%%%%%%%%%%%%%%%%%%%%%%%%%%%%%%%%%%%%%%%%%%%%%%%%%%%%%%%%%%%
%% FIGURE 
%%%%%%%%%%%%%%%%%%%%%%%%%%%%%%%%%%%%%%%%%%%%%%%%%%%%%%%%%%%%%%%%%%%%%%%%%%%%%%%

Algorithm \ref{alg:sat_attack} outlines the SAT attack methodology. A miter-like circuit is created using the locked netlist in the first step of the attack. In order to obtain the IC output $O\textsubscript{d}$, SAT solvers are used to derive a DIP $I\textsubscript{d}$ using the CNF formula of the miter. With the constraint on the \emph{diff} signal, the miter circuit is formulated as $L(I,K\textsubscript{A}) = L(I,K\textsubscript{B})$. In each iteration of the attack, a DIP $I\textsubscript{d}$ is obtained by identifying a satisfying assignment to the miter CNF. The output $O\textsubscript{d}$ of the functional IC is recorded, and the CNF formula is enriched with additional clauses based on the ($I\textsubscript{d}$, $O\textsubscript{d}$) input-output pair. When UNSAT is returned by the SAT solver, the attack is concluded. The formula $L(I,K\textsubscript{A})$ is solved in the final step of the SAT solver, resulting in the precise key K\textsubscript{C}. Repeatedly applying DIPs until the search space has been exhausted allows the attacker to obtain the correct key values.

\begin{algorithm}[htb]
\small
\caption{SAT attack algorithm \cite{eval_logic}}
\label{alg:sat_attack}
  \SetKwInOut{Input}{Input}
  \SetKwInOut{Output}{Output}
\Input{Locked netlist $L(I,K)$, functional IC $F(I)$}
\Output{Correct key $K\textsubscript{C}$}
\While{$I\textsubscript{d}= SAT(L(I,K\textsubscript{A}) = L(I,K\textsubscript{B}))$}{
$O\textsubscript{d} = F(I\textsubscript{d})$ \tcp*{Query the oracle}
$L(I,K\textsubscript{A}) = L(I,K\textsubscript{A}) \wedge (L(Id,K\textsubscript{A}) = O\textsubscript{d})$ \tcp*{Augment clauses}
$L(I,K\textsubscript{B}) = L(I,K\textsubscript{B}) \wedge (L(Id,K\textsubscript{B}) = O\textsubscript{d})$
}
$K\textsubscript{C}= SAT (L(I,K\textsubscript{A}))$;
\end{algorithm}

Oracle-less attacks are a set of techniques that do not require accessing an oracle or a functional chip. There are many types of oracle-less attacks, and most of them are based on structural analysis \cite{SIAL, SWEEP, SCOPE}. Although Oracle-guided attacks, such as the SAT attack, are applicable to some scenarios, they may not be effective in situations where the search space for the key bits is large. Recently, a few attacks have been proposed for reconfigurable-based obfuscations, which will be discussed in the following paragraphs.

In \cite{chowdhury2022predictive}, authors propose a ``predictive model attack'' that substitutes the exact logic mapped onto eFPGAs with a synthesizable predictive model, which attempts to replicate the behavior of the exact logic. This is an oracle-guided attack that relies on machine learning techniques to build a predictive model. %This attack is only applicable in the context of approximate computing, for instance, where hardware accelerators tolerate input errors to some extent. A widely assumed threat model in this context is that both the foundry and the end-user are considered untrusted. The main objective of an adversary is to reverse engineer with the intention of stealing its IPs, producing excess ICs, or potentially inserting advanced hardware trojans. To achieve this objective, the adversary is obligated to reconstruct the bitstream. The adversary possesses the GDSII file of the obfuscated design sent for fabrication. 
The adversary is a proficient IC designer with the expertise and tools necessary for understanding this layout representation. The threat model of \cite{chowdhury2022predictive} is given below:

\begin{itemize}
    \item Access to the inputs and outputs of an eFPGA can be facilitated by utilizing the scan-chain around the eFPGA. This approach is widely supported by all commercial eFPGA vendors.
    \item In the absence of a scan chain, a probing attack can be employed as an alternative method \cite{probing_attack}. Given the regular structure of the eFPGA, a probing attack would be relatively straightforward.
    \item Access to the CAD flow required for programming the eFPGA can be reasonably assumed, since very likely the eFPGA is licensed from a known vendor (e.g., Achronix, Menta, or Quicklogic).
\end{itemize}

The next proposed attacks are oracle-less attacks, namely the ``structural analysis attack'' and the ``composition analysis attack'' \cite{eASIC}. %The structural analysis attack aims to decrease the key search space for recovering the bitstream and it achieves that by eliminating impossible LUT configurations. On the other hand, the composition analysis attack aims to identify an obfuscated circuit by correlating its functionality to known circuits. 
Because \cite{eASIC} proposes a hybrid FPGA-ASIC solution, it is assumed that an attacker has access to a reversed engineered netlist where a portion of the logic is in the clear. The security of such style of reconfigurable-based obfuscation depends on how much information is exposed in this fully exposed portion of the logic. In \cite{Zain_TCAD}, a follow-up work, authors present the threat model for their attacks as follows:

\begin{itemize}
    
    \item The adversary may aim to learn the circuit's intent instead of recovering the entire IP (e.g., "Is this an AES cryptocore?"). To achieve this objective, the adversary does not need to recreate the correct bitstream, instead observations of a partially obfuscated logic give away hints about its intent.
    
    \item The adversary can recognize individual standard cells, therefore the gate-level netlist of the obfuscated circuit can be easily recovered following \cite{regds}.
    
    \item The ability of the adversary to identify reconfiguration pins \cite{SAT, yasin_logic_locking} allows for easy enumeration of all LUTs and also their programming order.
    
    \item The adversary can form clusters of standard cells existing in the static logic and transform them back into a LUT representation\footnote{In other words, perfect LUT reconstruction is assumed.}.
    
\end{itemize}

The last attack proposed in the literature is called ``break \& unroll attack''. It is capable of recovering the bitstream of eFPGA-based redaction schemes in a relatively short time even with the existence of hard cycles and large size keys \cite{eFPGA_sec_eval} and, interestingly, this contrasts the common perception of eFPGA-based redaction schemes being secure against oracle-guided attacks \cite{collini2022reconfigurable, mohan2021hardware, e-FPGA4, bhandari2021not, eASIC}. In their threat model, it is assumed that all ICs are sequential circuits, thus the attacker can access the always-present scan chain. The attacker has complete access to the obfuscated netlist and can obtain a functional circuit from the market as a black box, with the ability to derive correct outputs for given input vectors. 

In general, the proposed FPGA-inspired obfuscation techniques were mainly evaluated against SAT-based attacks which are logical in nature. Physical attacks, such as side-channel attacks, are barely assessed \cite{yang2018exploiting, kolhe2022lock, kolhe2021securing}. Overall, there is a lack of security analysis for evaluating the FPGA-inspired obfuscation techniques against specific attacks. For well-established attacks, the analyses performed so far still have a large margin for improvement.

\subsection{Attacks} \label{subsec:attacks}
FPGA-inspired obfuscation techniques, similar to the Logic Locking techniques, aim to protect IPs against different threats that appear at any phase in the globalized IC supply chain. The attack trend for the reconfigurable-based techniques is illustrated in Fig. \ref{fig:survey_trend}. There is an obvious lack of developed attacks against reconfigurable logic mechanisms---currently, there are a few developed attacks presented in  \cite{chowdhury2022predictive}, \cite{eASIC}, and \cite{eFPGA_sec_eval}. 

\begin {figure}[h!]
\centerline{\includegraphics[width=1.0\linewidth]{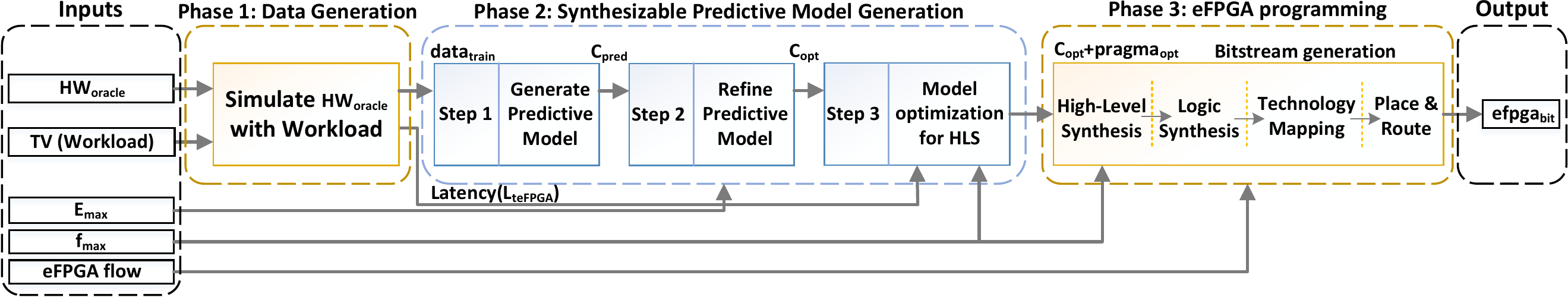}}
\caption{The proposed attack flow composed of three main phases, the predictive model attack (adapted from \cite{chowdhury2022predictive}).}
\label{fig:Predictive_model}
\end{figure}

The three main steps of a ``predictive model attack'' are illustrated in Fig. \ref{fig:Predictive_model}. In the first phase of the attack, the IC from the market is used to run applications that require the obfuscated hardware accelerator, and inputs and outputs of eFPGA are recorded for generating a predictive model. The latency of the obfuscated design is also recorded to ensure there are no timing issues when replacing the eFPGA portion with a predictive model. This text describes the second phase of the proposed attack, which involves searching for a predictive model that can be mapped in hardware on the eFPGA. The predictive model must fit within the eFPGA fabric, have outputs within the specified error threshold, and operate at the same frequency and latency as the original design. This phase is divided into three steps: model fitting, predictive model refinement, and automated MLP exploration. The output of this phase is a synthesizable C description of the predictive model that operates within the given constraints. The two predictive models considered are linear regression and multi-layer perceptron. The automated MLP explorer optimizes the MLP configuration to fit within the eFPGA and meet the error constraint. The predictive model generated in step 1 is further optimized in step 2 by obtaining the smallest possible model that operates within the specified error threshold. 

The third and final step in the process of generating a predictive model for High-Level Synthesis (HLS) involves finding the smallest possible implementation of the predictive model with a latency equal to that of the exact version extracted in the first phase. This is achieved by setting different synthesis options combinations for the optimized synthesizable predictive model and generating the smallest implementation with latency $L\textsubscript{efpga}$. The authors discuss the use of synthesis directives (pragmas) for synthesizing arrays, loops, and functions and how different combinations of these directives lead to a unique micro-architecture with specific area vs. latency trade-offs. The output of this stage is the pragma combination that leads to the smallest predictive model implementation (pragmaopt) and the newly optimized predictive model with exact latency as the exact obfuscated circuit (C\textsubscript{Copt}). The final phase of the process generates an eFPGA bitstream to configure an eFPGA with the predictive model. The process includes HLS, logic synthesis, technology mapping, place and route, and eFPGA bitstream generation. The target synthesis frequency should be set to the same frequency at which the exact obfuscated circuit works to enable the unlocking of every manufactured IC. Despite the cleverness of the attack, it does bare a major limitation since it only applies to approximate computing. It is therefore unlikely that the eFPGA-obfuscated logic would implement any form of hardware-based cryptography since it has to be deterministic.

The authors of \cite{Zain_TCAD} proposed two different attacks: ``structural analysis attack'' and the ``composition analysis attack''. %The majority of the LUTs are LUT\textsubscript{6} due to the packing algorithm executed during FPGA implementation.
It is argued that, by exploiting the static portion of the design, including the frequency of specific LUT masking patterns, an adversary can extract information and reduce the search space for the key that unlocks the design. They found that the combined number of unique masking patterns forms a set of 3376 elements, which reduces the global search space from $2^{64}$ to $3376=2^{11.72}$. They hypothesize that an attacker can exploit the frequency at which LUTs appear in a netlist to mount structural analysis attacks. The authors investigated this by analyzing the behavior of the frequency of masking patterns at different obfuscation levels. The results show that the adversary can estimate to some degree what masking patterns are the outliers. The results show that certain levels of obfuscation can either result in no correlation, a strong correlation to another circuit, or a correlation to itself. The study suggests that obfuscation can be targeted to confuse an adversary and shrink the key search space for obtaining the bitstream. 

However, the success of the attack depends on the adversary's ability to reconstruct LUTs from the static part, the availability of enough datapoints in the database, and the design having static parts to begin with (eFPGAs would not have static parts). Additionally, the frequency of masking patterns can serve as a template for comparing different designs, as the composition of the LUTs within a design would enable a powerful template-based attack.

In \cite{eFPGA_sec_eval}, the ``Break \& Unroll attack'' is proposed, which combines cycle breaking and unrolling to recover the bitstream of state-of-the-art eFPGA-based redaction schemes. The study highlights that the common perception that eFPGA-based redaction is secure against oracle-guided attacks is false and that additional research is required to secure eFPGA-based redaction schemes systematically. The overall flow of the Break \& Unroll attack is given in Algorithm \ref{alg:break_unroll}. The Break \& Unroll algorithm consists of two main stages: breaking cycles and unrolling remaining cycles. If the first stage fails to reveal the correct key, the second stage, unrolling, is employed to neutralize the effect of hard cycles. The breaking phase involves breaking all cycles and adding a non-cyclic condition as a new constraint to the obfuscated circuit. The unrolling phase conquers the weakness of the breaking phase by unrolling a single cycle at a time, choosing a single feedback and adding a copy of every gate to the circuit. After unrolling one cycle, a new version of the obfuscated circuit is created and set W must be updated each time in the body of the attack algorithm.

\begin{algorithm}[htb]
\small
  \caption{Break \& Unroll attack algorithm \cite{eFPGA_sec_eval}}
  \label{alg:break_unroll}
  \SetKwInOut{Input}{Input}
  \SetKwInOut{Output}{Output}
  
  \Input{Obfuscated circuit $g(x, k)$ and original function $f(x)$}
  \Output{Key vector $k^*$ such that $g(x, k^*) \equiv f(x)$}
  
  \While{(True)}{
    $W \leftarrow \text{SearchFeedbackSignals}(g(x,k))$ \\
    \tcp{ $W \leftarrow \{w_0, w_1, \dots, w_m\}$ }
    \For{$w_i \in W$}{
      $F(w_i, w_i') \leftarrow \text{BreakFeedback}(w_i)$ \\
      $NCCNF(k) \leftarrow \bigwedge\limits_{i=0}^{m} F(w_i, w_i')$ \\
      \tcp{ $NCCNF(k) \leftarrow \text{BreakFeedback}(w_0) \wedge \dots \wedge \text{BreakFeedback}(w_m)$ }
      $g(x, k) \leftarrow g(x,k) \wedge NCCNF(k)$ \\
    }
    $DIPset \leftarrow \emptyset$ \\
    \While{$\hat{x} \leftarrow \text{SAT}(g(x,k_1) \neq g(x,k_2))$}{
      \If{$\text{LoopDetected}(\hat{x}, DIPset)$}{
        $w \leftarrow \text{SelectFeedbackSignal}(W)$ \\
        $g(x, k) \leftarrow \text{NewCircuit}(w, g(x, k))$ \\
        \textbf{break} \\
      }
      $g(x, k_1) \leftarrow g(x, k_1) \wedge (g(\hat{x}, k_1) = f(\hat{x}))$ \\
      $g(x, k_2) \leftarrow g(x, k_2) \wedge (g(\hat{x}, k_2) = f(\hat{x}))$ \\
      $\text{Add}(\hat{x}, DIPset)$ \\
    }
    \If{$\neg\text{SAT}(g(x, k_1) \neq g(x, k_2))$}{
      \Return $k^* \leftarrow \text{SAT}(g(x, k_1))$ \\
    }
  }
\end{algorithm}

\begingroup
\setlength{\tabcolsep}{2.0pt} % Column spacing - Default value: 6pt
\renewcommand{\arraystretch}{1.1} % Row spacing  - Default value: 1
\begin{table} [htb]
\footnotesize \centering
\caption{Security comparisons of FPGA-inspired obfuscation techniques}
\label{tab:sec_comp}
\begin{tabular}{|p{0.4cm}|p{1.4cm}|p{1.2cm}|p{1.2cm}|p{1.2cm}|p{5.0cm}|} \hline
            % Header
            \multicolumn{2}{|c}{\multirow{2}{*}{\sc Obf. Technique}} &   \multicolumn{4}{|c|}{\sc Attack Resiliency}    \\ \cline{3-6}
            % Sub-header
	        \multicolumn{2}{|c|}{ } & {SA vs. AA} & {SAT} & {PSCA} & {Others} \\  \cline{1-6}
            \parbox[t]{4pt}{\multirow{17}{*}{\rotatebox[origin=c]{90}{\sc CMOS Tech}}}   
            % 1
            &  \cite{reconfigure_1} &  - &  {\textit {no}} & {\textbf{-}} & {\textbf{-}} \\ \cline{2-6}      
            % 2
            &  \cite{reconfigure_3} &  AA & {\textit {no}} & {\textbf{-}} & {Hardware-Based Code and Injection Attack} \\ \cline{2-6}     
            % 3
            &  \cite{lut1} &  AA & {\textit {yes}} & {\textbf{-}} & {\textbf{-}} \\ \cline{2-6}
            % 4
            &  \cite{Zain_TCAD} & AA &  {\textit {yes}} & {\textbf{-}} & {AppSAT, Removal Attack, Composition Attack, Structural Attack, and SCOPE} \\ \cline{2-6}
            % 5
            &  \cite{kolhe2022silicon} & AA & {\textit {yes}} & {\textbf{-}} & {Removal Attack} \\ \cline{2-6}
            % 6
            &  \cite{tomajoli2022alice} &  {\textbf{-}} &  {\textbf{-}} & {\textbf{-}} & {\textbf{-}} \\ \cline{2-6}     
            % 7
            &  \cite{e-FPGA4} & AA & {\textbf{-}} & {\textbf{-}} & {Icy-SAT} \\ \cline{2-6} 
            % 8
            &  \cite{mohan2021hardware} & AA & {\textit{yes}} & {\textbf{-}} & {\textbf{-}}       \\ \cline{2-6} 
            % 9
            &  \cite{e-FPGA1} &  AA &  {\textit{yes}} & {\textbf{-}} & {Brute Force Attack} \\ \cline{2-6}
            % & \cite{kim2015synthesizable} & {\textbf{-}} & {\textbf{-}} & {\textbf{-}} & {\textbf{-}} \\ \cline{2-6}  
            % &  \cite{gorbachov2020hardware} & {\textbf{-}} & {\textbf{-}} & {\textbf{-}} & Trojan Insertion \\ \cline{2-6}     
            & \cite{bhandari2021not} & AA & {\textit{yes}} & {\textbf{-}} & {Icy-SAT, CycSAT, and Be-SAT} \\ \cline{2-6}
            & \cite{chen2021area} & SA & {\textit{yes}} & {\textbf{-}} & Removal Attack \\ \cline{2-6}    
            & \cite{chowdhury2021enhancing} & AA & {\textit{yes}} & {\textbf{-}} & Removal Attack                  \\ \cline{2-6}     
            %& \cite{li2021janus} & SA & {\textit{yes}} & {\textbf{-}} & \makecell {Oracle-Less Structural Attack \\ FSM Reverse-Enginering \\ Functional FSM Separation \\ Partition Analysis \\ Functional Pruning Attack} \\ \cline{2-6}
            & \cite{lut5} & AA & {\textit{yes}} & {\textbf{-}} & {Removal Attack, Scan-Based Attack, ATPG-Based Attack, Approximate Attack, and SMT-Based Attack} \\ \hline 
            & \cite{e-FPGA3} & SA & {\textit{yes}} & {\textbf{-}} &  {Brute Force Attack} \\ \cline{2-6}          
	    	\parbox[t]{4pt}{\multirow{9}{*}{\rotatebox[origin=c]{90}{\sc Emerging Tech}}} &  \cite{patnaik2018advancing} & AA &  {\textit{yes}} & {\textbf{-}} & {Double DIP} \\ \cline{2-6}
            %&  \cite{rangarajan2020opening} & AA &  {\textbf{-}}     & {\textbf{-}}    & \makecell {AppSAT \\HACKTEST (Oracle Less Attack)} \\ \cline{2-6} 
	    	&  \cite{kolhe2021securing} &  AA &  {\textit{yes}} & {\textit{yes}} & {AppSAT, Removal Attack, Scan and Shift-Based Attack} \\ \cline{2-6}
	    	& \cite{kolhe2022lock} & AA & {\textit{yes}} & {\textit{yes}} & {Removal Attack, Scan and Shift-Based Attack} \\ \cline{2-6} &  \cite{winograd2016hybrid} &  SA &  {\textit{no}} & {\textbf{-}} & {Brute Force Attack, Machine Learning-Based Attack} \\ \cline{2-6}
	    	&  \cite{yang2018exploiting} &  SA &  {\textbf{-}} & {\textbf{-}} & {Brute Force Attack, Side-Channel Based Attack, Testing-Based Attack, Circuit Partition-Based Attack} \\ \cline{2-6}
	    	&  \cite{lut2} &  AA &  {\textit{yes}} & {\textit{yes}} & {\textbf{-}} \\ \cline{2-6}
	    	& \cite{lut3} &  AA &  {\textit{yes}} & {\textbf{-}} & {AppSAT, Removal Attack, Scan, and Shift-Based Attack} \\ \cline{2-6}
	    	& \cite{attaran2018static} & {\textbf{-}} &  {\textit{no}} & {\textbf{-}} & {\textbf{-}} \\ \hline
\end{tabular}
\begin{tablenotes}
      \centering
      \item  \footnotesize SA and AA are abbreviations for Security Analysis and Applied Attack.  
    \end{tablenotes}
%\vspace{-2mm}
\end{table}
\endgroup

\subsubsection{State Space Reduction} \label{subsubsec:space_reduction}
\begin{figure}[tb]
\centering 
\includegraphics[width=0.48\linewidth]{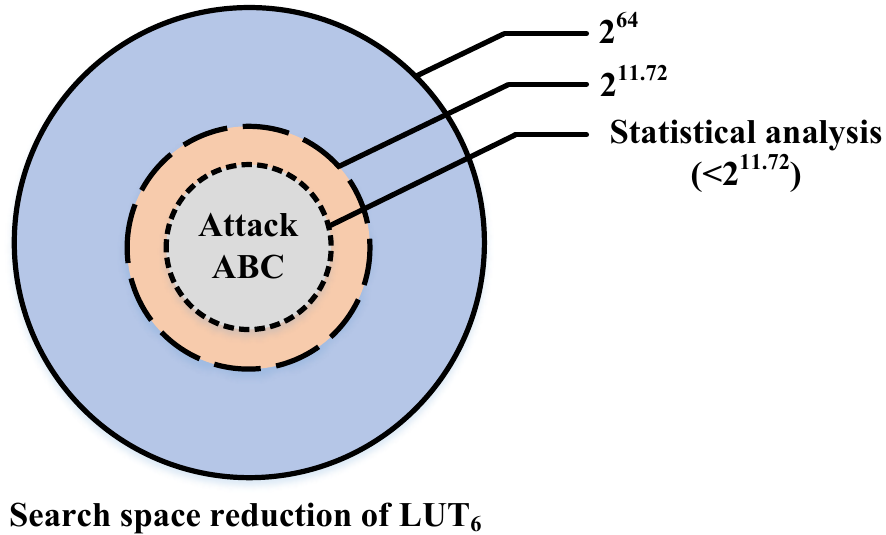}
\caption{The search space of LUT\textsubscript{6} as it shrinks with different attacks (adapted from \cite{eASIC}).}
\vspace{-10pt}
\label{fig:search_space}
\end{figure}

The large search space of the bitstream makes reconfigurable-based obfuscation techniques inherently resilient to SAT attacks. However, a smart adversary could reduce the state space of a bitstream, as discussed in \cite{eASIC}. The authors explain that the synthesis tool may not explore all $2^n$ possible configuration patterns for n-input LUTs. Moreover, the adversary can exploit the frequency distribution of specific LUTs that are heavily preferred by synthesis engines. With the frequency characteristics of the LUT combined with statistical analysis, the search space can be reduced significantly. The significance of minimizing the search space lies in its criticality since it allows the adversary to identify the regular patterns and potential information inside the circuitry. Let us say the largest LUT in the design is a 6-input LUT. In this case, the search space is explored with the size of $2^{64}$ making patterns as illustrated in Fig. \ref{fig:search_space}. The authors in \cite{eASIC} considered dozens of designs of varying complexity, size, and functionality. As a result, a total of $X = 2^{11.72}$ unique masking patterns were generated reducing the global search space significantly, as illustrated in Fig. \ref{fig:search_space}. By incorporating synthesis hints along with statistical analysis, the adversary effectively narrows down the search space to another subsequent level. Furthermore, at that point, the adversary can apply his/her attack "ABC", whether it is a guided SAT-based attack or another clever attack. These considerations remain to be explored in future work.

\subsection{Comparisons} \label{sec:comparisons}
Table \ref{tab:sec_comp} presents a list of reconfigurable-based obfuscation techniques, categorized according to the technology used in the implementations, as discussed in Section \ref{sec:Principles_com}. The prohibitive costs of reconfigurable-based obfuscation techniques could make them impractical despite their attack resiliency. As an attempt to decrease the overheads without compromising security, researchers have explored emerging non-volatile memory technologies for implementing reconfigurable logic. In addition, NVM technology can be used as a replacement for a tamper-proof memory \emph{if} they lose their content upon invasive reverse engineering attempts \cite{lut3}. Despite being promising technologies, these technologies are still under development and have not been widely adopted. For a detailed discussion, we refer our readers to \cite{nvm_future}.

As indicated in Table \ref{tab:sec_comp}, some authors solely employed security analysis (SA) to assess the effectiveness of their defense techniques. If authors effectively attempted to attack their own defences, we marked them as applied attacks (AA). Here, we make a clear argument that both SA and AA have merits, but SA can be easily misinterpreted to suggest the techniques are more secure than they actually are. For instance, a classical SA discussion is the enumeration of the adversarial search space. But even large search spaces can be broken and this can only be shown through AA. 

Furthermore, the resilience of the techniques was evaluated mainly against SAT-based attacks, as listed in the fourth column of Table \ref{tab:sec_comp}. However, it should be noted that most of the techniques have not been validated against power side-channel attacks (PSCAs), especially in the case of CMOS implementations. The authors of \cite{yang2018exploiting} and \cite{e-FPGA3} discussed the resiliency of their techniques against testing-based attacks, circuit partition-based attacks, brute force attacks, SAT-attacks, and side-channel attacks. The column ``Others'' in the table shows the resilience of a given technique against other applied attacks. We have observed this trend over and over with Logic Locking mechanisms that were considered secure -- some even proven to be secure -- falling prey to \emph{simple} attacks that were simply not initially considered.

%%%%%%%%%%%%%%%%%%%%%%%%%%%%%%%%%%%%%%%%%%%%%%%%%%%%%%%%%%%%%%%%%%%%%%%%%%%%%%%
%% FIGURE 
%%%%%%%%%%%%%%%%%%%%%%%%%%%%%%%%%%%%%%%%%%%%%%%%%%%%%%%%%%%%%%%%%%%%%%%%%%%%%%%
\begin {figure}[h!]
\begin{center}
\begin{tikzpicture}
\begin{axis}[
xtick=0,
legend columns=3, 
legend style={font=\small, legend cell align=right}, 
legend pos=north west,
xlabel={Techniques},%
ylabel={Ratio}]% 
\addplot[color= blue,  mark= Mercedes star flipped, mark size= 2, only marks,]  
coordinates{
(2, 0.313)       %\cite{patnaik2018advancing}
(2, 0.651)       %\cite{patnaik2018advancing}
(2, 1.25)        %\cite{patnaik2018advancing}
};
\addlegendentry{\cite{patnaik2018advancing}}
\addplot[color= blue,  mark= pentagon*, mark size= 2, only marks,]  
coordinates{
(3, 1.109)      % \cite{yang2018exploiting}
(3, 18.191)     % \cite{yang2018exploiting}
(3, 70.844)     % \cite{yang2018exploiting}
};
\addlegendentry{\cite{yang2018exploiting}}
\addplot[color= red,  mark= pentagon, mark size= 2, only marks,]  
coordinates{
(4, 3.119)       % \cite{e-FPGA4}
(4, 12.504)      % \cite{e-FPGA4}
(4, 20.035)      % \cite{e-FPGA4}
};
\addlegendentry{\cite{e-FPGA4}}
\addplot[color= blue,  mark= triangle*, mark size= 2, only marks,]  
coordinates{
(5, 0.0118)      % \cite{eASIC}
};
\addlegendentry{\cite{eASIC}}
\addplot[color= blue,  mark= diamond*, mark size= 2, only marks,]  
coordinates{
(6, 0.008)        % \cite{kolhe2022silicon}
(6, 9.056)        % \cite{kolhe2022silicon}
(6, 55.038)       % \cite{kolhe2022silicon}
};
\addlegendentry{\cite{kolhe2022silicon}}
\addplot[color= blue,  mark= Mercedes star, mark size= 2, only marks,]  
coordinates{
(7, 51.988)      % \cite{lut5}
(7, 54.876)      % \cite{lut5}
(7, 57.764)      % \cite{lut5}
};
\addlegendentry{\cite{lut5}}
\addplot[color= blue,  mark= diamond, mark size= 2, only marks,]  
coordinates{
(8, 2.756)      % \cite{lut2}
(8, 22.518)     % \cite{lut2}
(8, 63.625)     % \cite{lut2}
};
\addlegendentry{\cite{lut2}}
\addplot[color= blue,  mark= o, mark size= 2, only marks,]  
coordinates{
(9,1.28)        %\cite{lut3}
(9,7.966)       %\cite{lut3}
(9,18.179)      %\cite{lut3}
};
\addlegendentry{\cite{lut3}}
\addplot[color= red,  mark= heart, mark size= 2, only marks,]  
coordinates{
(10, 3.482)       % \cite{chowdhury2021enhancing}
(10, 56.619)      % \cite{chowdhury2021enhancing}
(10, 187.991)     % \cite{chowdhury2021enhancing}
};
\addlegendentry{\cite{chowdhury2021enhancing}}
\addplot[color= blue,  mark= x, mark size= 2, only marks,]  
coordinates{
(11,39.046)      %\cite{kolhe2021securing}
(11,71.416)      %\cite{kolhe2021securing}
(11,86.317)      %\cite{kolhe2021securing}
};
\addlegendentry{\cite{kolhe2021securing}}
\addplot[color= blue,  mark= triangle, mark size= 2, only marks,]  
coordinates{
(12, 0.225)       %\cite{rangarajan2020opening}
(12, 1.863)       %\cite{rangarajan2020opening}
(12, 5.163)       %\cite{rangarajan2020opening}
};
\addlegendentry{\cite{rangarajan2020opening}}
\addplot[color= blue,  mark= star, mark size= 2, only marks,]  
coordinates{
(13, 5.721)       % \cite{bhandari2021not}
(13, 7.882)       % \cite{bhandari2021not}
(13, 9.574)       % \cite{bhandari2021not}
};
\addlegendentry{\cite{bhandari2021not}}
\addplot[color= blue,  mark= halfcircle*, mark size= 2, only marks,]  
coordinates{
(14, 4.528)       % \cite{eFPGA_sec_eval}
(14, 6.629)       % \cite{eFPGA_sec_eval}
(14, 8.657)       % \cite{eFPGA_sec_eval}
};
\addlegendentry{\cite{eFPGA_sec_eval}}
\end{axis}
\end{tikzpicture}
\caption{Number of gates  to the length of bitstream ratio in several techniques. The red color indicates that corresponding data was estimated, whereas the blue color indicates that the data was given in related references.}
\label{fig: ratio}
\end{center}
\end{figure}
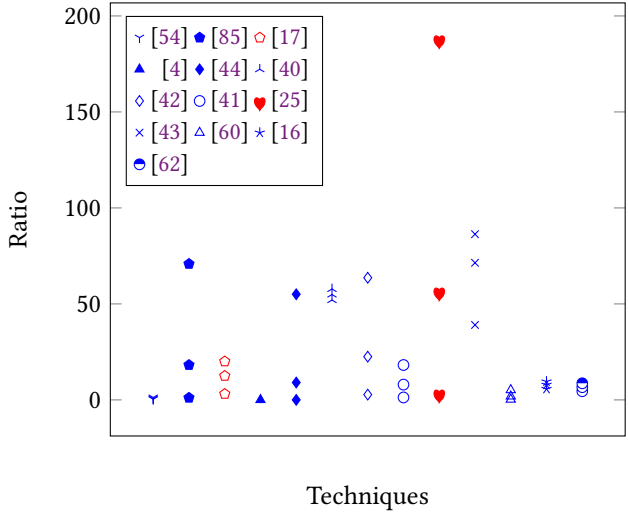
%%%%%%%%%%%%%%%%%%%%%%%%%%%%%%%%%%%%%%%%%%%%%%%%%%%%%%%%%%%%%%%%%%%%%%%%%%%%%%%
%%%%%%%%%%%%%%%%%%%%%%%%%%%%%%%%%%%%%%%%%%%%%%%%%%%%%%%%%%%%%%%%%%%%%%%%%%%%%%%

One of the major challenges is defining a security metric for reconfigurable-based obfuscation techniques. Currently, there is no specific or common security metric being adopted. To fairly assess the security of reconfigurable-based obfuscation techniques in this study, we adopted a common criterion: number of gates divided by the length of bitstream. Here the term ``gates" refers to logic gates in the same was as in logic synthesis, meaning that a gate is a standard cell. For instance, if an author considered that a circuit with 1000 gates was adequately obfuscated with a bitstream of 128 bits, this circuit would have a metric of $7.81$. Authors of defensive techniques are interested in maximizing the metric, such that a high number of gates can be protected with a small number of configuration/programming bits.

Often, authors report the area of obfuscated design, and from that, we can estimate the number of gates from the area using information about the technology used for implementation. Fig. \ref{fig: ratio} compares the gates count to the bitstream size ratio of benchmark circuits to be obfuscated as a security metric. We calculated the minimum, maximum, and average values of the ratio in most of the techniques\footnote{Not all surveyed papers provide enough information for this comparison, unfortunately.} considering all benchmarks and bitstream sizes used originally by the developers for evaluating their techniques. As a rule of thumb, a large ratio is an indication of lower and costly security \cite{lut1, patnaik2018advancing, e-FPGA3}. However, there are exceptions to this rule \cite{lut1}. In addition, optimizing a design for desirable PPA overheads and security would complicate the security-bitstream size relation \cite{bhandari2021not}. For some techniques, calculating the ratio was straightforward as the gates count and the size of bitstream were provided by the authors \cite{eASIC, kolhe2022silicon, lut3}. However, for other techniques, the ratio either could not be calculated due to missing values \cite{reconfigure_1}, \cite{lut1} or estimations were required for calculating it. For example, in \cite{e-FPGA4}, we needed to turn the area of a given circuit into the gate equivalent (GE) form to estimate the circuit gate counts.

We noticed two contradictory solutions in the eFPGA context: First, in many techniques, obfuscating nearly 10\%, 20\%, or 30\% of a circuit gates count could suffice to ensure security (i.e., SAT attacks would run out of time) \cite {reconfigure_1}, \cite{lut1}, \cite{patnaik2018advancing}, \cite{e-FPGA3}. Second, in \cite{eASIC}, a much higher obfuscation rate is required for protecting the bitstream and design intent, e.g., obfuscation rates higher than 86\% are recommended. As a result, the first solution would result in a large ratio when compared to the second solution as this is evident in Fig. \ref{fig: ratio}. Yet, the objectives of the techniques are different, and so are the underlying threat models. 

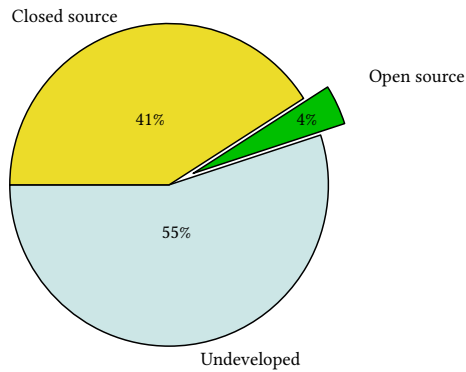
\begin {figure}[h!]
\begin{center}
\resizebox{0.45\textwidth}{!}{
\begin{tikzpicture}
\pie[rotate=180, color={teal!20, green!75!black, yellow!90!black}, explode = { 0, .5, 0}]{55/Undeveloped, 4/Open source, 41/Closed source}
% selecting different colors for each slice
% explode option makes a slice pop
 \end{tikzpicture}}
\caption{Reconfigurable obfuscation techniques categorization based on developed/undeveloped CAD tools. The developed tools are split into open source/closed source.}
\label{fig: DTools}
\end{center}
\end{figure}

We have also surveyed the CAD tools developed by the authors of obfuscation techniques. In 41\% of the proposed techniques, researchers developed custom tools to implement their approaches, whereas, in other techniques, researchers only relied on the standard CAD tools for implementing their approaches. The developed tools are mostly closed source and are claimed to be compatible with standard CAD flow as illustrated in Fig. \ref{fig: DTools}. 

%%%%%%%%%%%%%%%%%%%%%%%%%%%%%%%%%%%%%%%%%%%%%%%%%%%%%%%%%%%%%%%%%%%%%%%%%%%%%%%
%%%%%%%%%%%%%%%%%%%%%%%%%%%%%%%%%%%%%%%%%%%%%%%%%%%%%%%%%%%%%%%%%%%%%%%%%%%%%%%
%%%%%%%%%%%%%%%%%%%%%%%%%%%%%%%%%%%%%%%%%%%%%%%%%%%%%%%%%%%%%%%%%%%%%%%%%%%%%%%
%%%%%%%%%%%%%%%%%%%%%%%%%%%%%%%%%%%%%%%%%%%%%%%%%%%%%%%%%%%%%%%%%%%%%%%%%%%%%%%
%% FIGURE 
%%%%%%%%%%%%%%%%%%%%%%%%%%%%%%%%%%%%%%%%%%%%%%%%%%%%%%%%%%%%%%%%%%%%%%%%%%%%%%%

%%%%%%%%%%%%%%%%%%%%%%%%%%%%%%%%%%%%%%%%%%%%%%%%%%%%%%%%%%%%%%%%%%%%%%%%%%%%%%%

\section{Discussion} \label{sec:discussion}
In this section, we compare reconfigurable-based obfuscation and Logic Locking techniques and emphasize the lack of tape-outs demonstrating reconfigurable-based obfuscation techniques in silicon.

\subsection{Reconfigurable-based obfuscation vs. Logic Locking} \label{sec:obfuscation_locking}
The Logic Locking concept has been around for more than a decade. In recent years, a cat-and-mouse game was established between developing and attacking Logic Locking techniques. This area of research has progressed rapidly, resulting in dozens of defense techniques and published attack strategies. The research line of reconfigurable-based obfuscation is not a new one, but it has received far less attention than Logic Locking. Both Logic Locking and reconfigurable-based obfuscation techniques share a common objective of protecting IP against supply-chain attacks. 

\begin {figure}[ht]
\centerline{\includegraphics[width=0.65\linewidth]{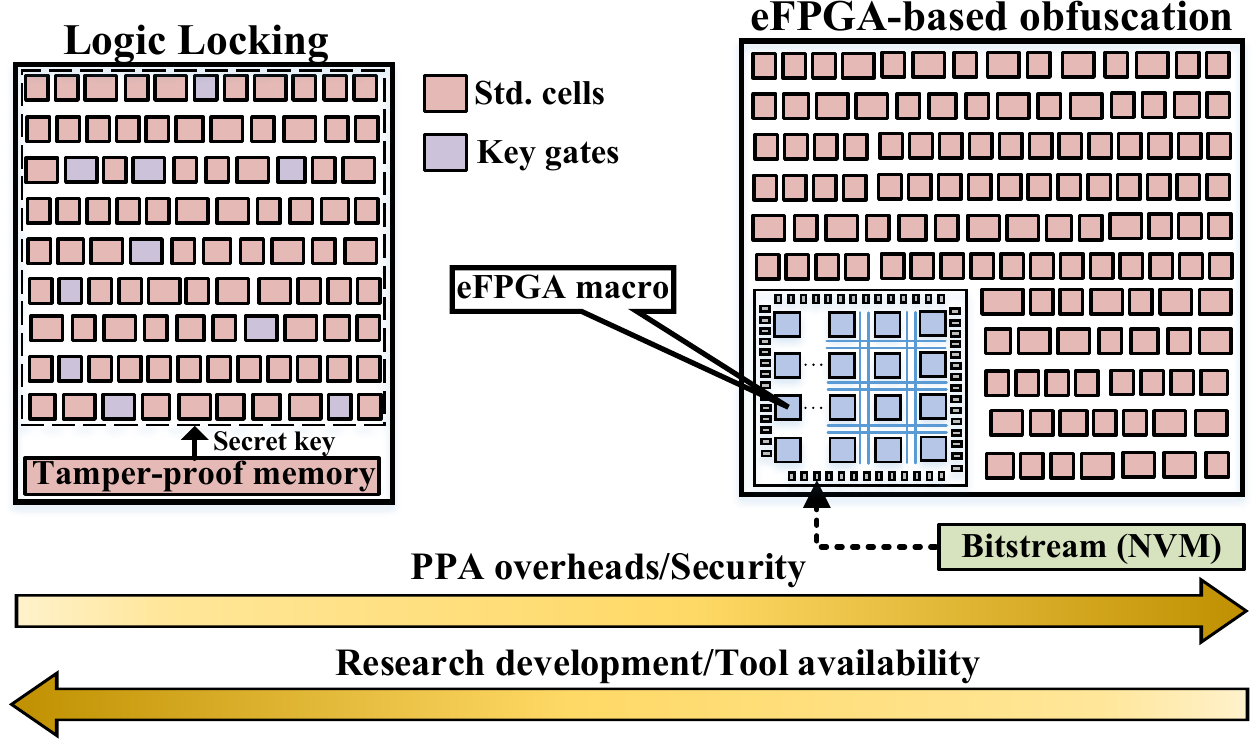}}
\caption{Conceptual diagram of Logic Locking vs. eFPGA redaction}
\label{fig:ReconfigL_LL}
\end{figure}

Initially, Logic Locking techniques received more attention due to their practicality -- the process of inserting XOR/XNOR gates in a netlist can be easily scripted. However, advances in Logic Locking have coincided with emerging powerful attacks that have compromised design security. Recently, reconfigurable-based obfuscation has received more attention due to its high resiliency against attacks. However, this has raised questions about the security versus PPA trade-offs. As a result, researchers have been proposing solutions where the reconfigurable part is as small as possible \cite{lut5}. Fig. \ref{fig:ReconfigL_LL} illustrates the conceptual difference between the Logic locking approach, shown on the left, that involves adding key gates to the original design, whereas the reconfigurable-based obfuscation approach, shown on the right, leverages reconfigurable logic elements. Logic Locking requires the correct configuration of the secret key while reconfigurable-based obfuscation techniques rely on loading the correct bitstream. These approaches are somewhat analogous and it is for this reason that attacks from Logic Locking domain can also be applied to reconfigurable solutions.

In Logic Locking, the locking mechanism is embedded in the netlist of design so the design is locked behind the secret key. This procedure could subject the design to several attacks, e.g., identifying and removing the lock, tampering with the lock, or identifying the secret key, compromising the secured design. The adversary can access the entire design consisting of the original intellectual property combined with the key gates. On the other hand, a selected portion of the design is hidden in reconfigurable-based obfuscation, meaning that a portion of the design is exposed to the adversary. This technique, which does not expose the entire design to the adversary, appears to offer a higher potential for securing ICs against supply-chain attacks.

In Logic Locking, circuit designers utilize the conventional CAD design flow. Reconfigurable-based obfuscation lacks a CAD tool flow that can support combined logic synthesis, timing analysis, and optimization of a mixed ASIC and FPGA-like design, unlike Logic Locking \cite{mohan2021hardware}. The right panel of Figure \ref{fig:ReconfigL_LL} shows an example of reconfigurable-based obfuscation, such as the eFPGA redaction technique and it employs custom tools that require significant effort to build. Reconfigurable-based obfuscation incurs higher security and overhead costs compared to Logic Locking, as shown by the arrows pointing to the right in Figure \ref{fig:ReconfigL_LL}. It appears that Logic Locking techniques are more mature as well as more readily available tools. To summarize, reconfigurable obfuscation presents promising opportunities for securing digital integrated circuits against supply chain attacks, though many challenges remain.

\subsection{Lack of Silicon Validation} \label{sec:tapeout}
There is a severe lack of frameworks to validate the security and functionality of obfuscated designs, as correctly highlighted in \cite{kolhe2022silicon, split_1}. Additionally, there is a minor effort toward silicon validation of the techniques as a proof of concept. For example, only techniques presented in \cite{kolhe2022silicon} and \cite{e-FPGA3} were validated on silicon. Silicon validations of the techniques would provide an impetus to advance reconfigurable-based obfuscation as a promising solution for protecting digital ICs.

\subsection{Future trends and Challenges} \label{sec:future_trends}
During our survey, an effort was made to present the findings of numerous research papers in a clear and unbiased manner. Yet, it is as evident as ever that the hardware security community lacks a unified benchmark suite and/or a common set of criteria. Frequently, researchers employ benchmark suites that enjoy popularity within the test community but are irrelevant to security. For instance, the ISCAS’85 suite lacks crypto cores and other real applications that are the cornerstone of the evaluation in this field. Furthermore, it is our belief that employing circuits that more accurately reflect current IC design practices, where IPs frequently comprise millions of gates and ICs house billions of transistors, would be highly beneficial for the community.

Within the research community, no standard criteria for assessing the security of reconfigurable-based obfuscation techniques is observed. Although it is confirmed that such techniques are resilient to various state-of-the-art attacks, the need for commonly accepted criteria to evaluate their security is an urgent need. While we utilized a ratio of gates to bitstream size, which is a reasonable but simple metric, it is noteworthy that for eFPGA estimations need to be converted into the number of gates. Concerning eFPGA macros available on the market, few companies have products available, including Achronix \cite{achronix}, Menta \cite{menta}, and Quicklogic \cite{quicklogic}. In tandem, there is also a discernible trend in the realm of ASIC design companies, such as AMD \cite{amdacquiredxilinx} and Intel \cite{intelacquiredaltera}, which both have acquired FPGA technology. This strategic move has enabled these entities to benefit from both ASIC and FPGA domains. With the convergence of these two technologies, there exists a possibility of integrating reconfigurable logic as an integral aspect of commercial production for security purposes. Today, the driver behind ASIC-FPGA hybrid solutions are design metrics, not security metrics. %For instance, Intel's recent product launch of ``eASIC'' has resulted in the successful commercialization of three different varieties \cite{intel_eASIC}. Employing this practice facilitates the production of secure ASICs by silicon companies.

It is also worth discussing the threat models proposed so far. The authors in \cite{Zain_TCAD} have established a robust threat model. Defining the capabilities of an attacker remains a complex task, requiring an understanding of their motivations, technical skills, and resource availability. Underestimating the attacker may result in ineffective defense strategies, whereas overestimating them could lead to unnecessary PPA overheads due to convoluted defense strategies. This challenge extends to reconfigurable-based obfuscation techniques and any other obfuscation-promoting approaches. 

Another consideration in terms of attack is whether an attacker can leverage a partially recovered netlist. For instance, in a design that employs multiple instances of the same block, recovering one block correctly may enable the attacker to recover all other instances of the same block through a visual inspection of their structure \cite{our_hierarchy}. This line of thinking is also applicable to datapaths and certain cryptographic structures that exhibit regularity. Consequently, a functional analysis of the recovered netlist can be combined with existing attacks to enhance correctly guessed connections.% In \cite{Zain_TCAD}, the authors presented two distinct attacks: the ``structural analysis attack'' and the ``composition analysis attack''. The authors exploit the regularity and frequency of LUTs to reduce the search space, while the latter utilizes the regularity and known circuit database to infer the design's functionality.

%Additionally, it is worth noting that at least one machine learning-based work has defined a distinct threat model. According to \cite{chowdhury2022predictive}, access to CAD flows is necessary to program the eFPGA, and a tool flow from a third-party vendor is also required. In the threat model, this assumption is irrelevant since the bitstream is independent of the CAD and tool flow. It makes no sense to contact third-party vendors for information about programming the eFPGA since it does not require information from these sources. Alternatively, an adversary could benefit from the exposed portions. A malicious adversary may also be able to obtain the necessary programming information if they have access to eFPGA's datasheet. The adversary could benefit from SAT-attacks along with the other attacks, i.e., testing-based attacks, circuit partition-based attacks, brute force attacks, probing attacks, and side-channel attacks. Furthermore, adversaries could exploit hints and heuristics generated by EDA tools and eFPGA vendors' tool flows. Clever reverse engineering techniques can capture these logical hints.

Figure \ref{fig: DTools} clarifies that developing a custom tool is almost a mandatory part of the obfuscation process: 45\% of the researchers have developed their custom tool, and 41\% of them are closed source. This practice is a barrier to the research community and there must be an initiate to make the tools open-source for academic use. Future research could (and should) explore how these tools, along with reverse engineering and other methods mentioned previously, can be utilized to break the security of reconfigurable-based obfuscation. Logic Locking has become increasingly susceptible to SAT attacks. There have been many attacks and measures against these attacks, so the Logic Locking concept continues to evolve. In spite of the proposal of SAT-hard solutions for Logic Locking, perhaps reconfigurable-based obfuscation techniques will eventually take over due to its inherit security against SAT-attacks. It is predicted that eFPGA-based obfuscation will gradually surpass Logic Locking in the coming years due to the fact that eFPGAs offers reliable security. Researchers in this field are likely to benefit from open-source macro generators for eFPGAs since only commercial solutions exist in the market. In the near future, there will probably be many publications based on this area of research.

%It is noteworthy that several of the works reviewed in this survey have yet to demonstrate their approach in silicon, as outlined in Section \ref{sec:tapeout}. As a community, we should endeavor to validate our approaches in silicon whenever possible. However, it is worth noting that reconfigurable-based techniques are becoming increasingly mature and are being adopted commercially, such as eASIC from Intel. This is likely a contributing factor to the relatively low percentage of works included in this survey that have validated their techniques in silicon.

%%%%%%%%%%%%%%%%%%%%%%%%%%%%%%%%%%%%%%%%%%%%%%%%%%%%%%%%%%%%%%%%%%%%%%%%%%%%%%%
%% Conclusions
%%%%%%%%%%%%%%%%%%%%%%%%%%%%%%%%%%%%%%%%%%%%%%%%%%%%%%%%%%%%%%%%%%%%%%%%%%%%%%%

\section{Conclusion} \label{sec:conclusions}
%The semiconductor industry is growing quickly and is now the backbone of the current computing system that propels the globalization of the IC supply chain. This also increases the associated hardware threats, especially on hardware protection and IP piracy. As a result, during the past ten years, a lot of attention has been paid to the research on hardware IP protection using reconfigurable-based logic obfuscation. 

Our study revealed a significant variation in the approaches to the reconfigurable-based obfuscation techniques among the surveyed works. However, we were able to classify the studies, providing a clear demonstration of the many interpretations of the technique, its attacks, and defenses. %First, we present the complete literature study on all defensive techniques by dissecting reconfigurable-based obfuscation into four different categories of abstraction including the technology used, element type, and IP type. Our survey focused on publications that appear in prestigious venues, ensuring that we assessed the most significant studies on reconfigurable-based obfuscation techniques. 
The study results were compiled to provide essential features, metrics, and performance results. %They were presented in a manner that illustrates the current state of the technique, our classification of these techniques will be helpful for future researchers to contextualize their own approaches to understand better the reconfigurable-based obfuscation techniques as illustrated in Fig. \ref{fig:classification}. 
Table \ref{tab:sec_3} provides valuable insights into the impact of different obfuscation methods on key design metrics and can help designers to select an appropriate obfuscation technique based on specific design requirements and constraints. Design and fabrication are becoming more complex with emerging technologies. The research community needs to address the deficiency in the reconfigurable elements of these technologies, even though their developers claim these elements display high-reliability and low-area requirement. 

%Overall, the majority of reconfigurable-based techniques leverage either a reconfigurable element or the eFPGA redaction. Despite the satisfactory security level provided by reconfigurable-based techniques, which offer a reasonable guarantee against supply chain attacks, the attacks on reconfigurable-based obfuscation techniques are not yet fully developed. However, reconfigurable-based obfuscation techniques require validation for state-of-the-art attacks even though they are SAT-resistant. 
The evaluation metrics and benchmark suites used varied widely, making direct comparisons challenging and, in some cases, impossible. The ``ratio of gates to bitstream length'' is a straightforward yet sound criterion for assessing the security of reconfigurable-based obfuscation techniques. %Table \ref{tab:sec_comp} summarizes the security analysis performed by different authors for their proposed techniques. 
Based on Fig. \ref{fig: ratio}, most techniques fall within 0-50 range for this ratio, indicating varying degrees of security. %The trend illustrated in Fig. \ref{fig:survey_trend} demonstrates that these techniques provide encouraging results for security, and no attack has yet been found that can completely break the security. 
Currently, there are a few developed attacks presented in  \cite{chowdhury2022predictive}, \cite{eASIC}, and \cite{eFPGA_sec_eval}, and no attack has yet been found that can achieve perfect success. By leveraging state space reduction along with other attack strategies, likely further advances in attack results can be achieved. 

It is an urgent need to conduct a comprehensive evaluation of critical security aspects to ensure the viability of the many proposed techniques. In the meantime, the research community must also validate the techniques on silicon to quickly eliminate approaches that are not viable, either from a technology standpoint of from a PPA overhead perspective. %It is imperative that design houses adopt these security measures to make them more practical, according to the current forecast. 

\bibliographystyle{ACM-Reference-Format}
\bibliography{sample-base}

%\appendix


\end{document}